\documentclass[a4paper,10pt]{article}

\usepackage[top=1in,bottom=1in,left=0.75in,right=0.75in]{geometry}
\usepackage [dvips]{graphicx}
\usepackage{amssymb,color,fancyhdr}

              \title{Two dimensional outflows for cellular automata\\ with shuffle updates}
   \def\shorttitle{2D outflows for cellular automata with shuffle updates}
           \author{Chikashi Arita$^1$, 
                     Julien Cividini$^2$\footnote{
                         Now in Physics of Complex Systems, Weizmann Institute of Science, Rehovot 76100, Israel}, 
                     C\'ecile Appert-Rolland$^2$ } 
\def\shortauthor{C Arita, J Cividini, C Appert-Rolland} 
     \def\address{1 Theoretical Physics,  Saarland University,  66041 Saarbr\"ucken, Germany\\ 
                     2  Laboratory of Theoretical Physics, CNRS (UMR 8627) and\\
                          University Paris-Sud, Building 210 91405 Orsay Cedex, France}
         \def\abst{In this paper, we explore the two-dimensional behavior of cellular automata with shuffle updates. As a test case, we consider the evacuation of a square room by pedestrians modeled by a cellular automaton model with a static floor field. Shuffle updates are characterized by a variable associated to each particle and called \textit{phase}, that can be interpreted as the phase in the step cycle in the frame of pedestrian flows.  Here we also   introduce a dynamics  for these phases, in order to modify the properties of the model. We  investigate  in particular  the crossover between low- and high-density regimes  that occurs  when the density of pedestrians increases, the dependency of the outflow in the strength of the floor field, and the shape of the queue in front of the exit. Eventually we discuss the relevance of these results for pedestrians.
}
              \date{ }

\makeatletter

\def\@maketitle{ 
\begin{center} 
  \let \footnote \thanks
    {\LARGE\linespread{1.2}\selectfont\textbf{\@title}\par}  \vskip 10mm 
    {\Large \@author}           \vskip 5mm  
    {\address}   \vskip 5mm 
    {\large \@date}              \vskip 5mm 
    \textbf{Abstract}     \end{center}
  \begin{quote} \abst \end{quote}   \vskip 5mm  
\noindent\makebox[\linewidth]{\rule{\textwidth}{0.5pt}}}

\makeatother

\fancypagestyle{titlepage}{
 \lhead{}  \chead{}  \rhead{}   
 \lfoot{}  \cfoot{\thepage}  \rfoot{}    }

\begin{document}

\maketitle

\thispagestyle{titlepage}

\section{Introduction}

Modeling traffic systems has been a problem of growing interest in statistical physics during the last decade~\cite{schadschneider2002a, helbing2001b,chowdhury_s_s2000,chou_m_z2011,appert-rolland_e_s2015}.  Indeed, these problems usually involve a large number of interacting driven agents that exhibit collective effects, and therefore belong to the realm of nonequilibrium statistical mechanics. Some examples of these collective effects are given by the spontaneous congestion of a highway~\cite{nagel_s1992,knospe2002b} or the pattern formation observed in counter-propagating lanes of pedestrians~\cite{hoogendoorn_d2003b,daamen_h2003a, moussaid2012} or at intersections~\cite{hoogendoorn_d2003b,daamen_h2003a, plaue2011, zhang_s2014}. Such effects can often be reproduced in minimal models inspired by statistical physics~\cite{barlovic1998,appert_s2001,burstedde2001a,burstedde2001b, kirchner2003a, schadschneider_s2011,cividini_a_h2013,masuda_n_s2014}. Many of these models belong to the class of cellular automata. In these models, the moving agents (cars, pedestrians, molecular motors, etc) are usually modeled as particles hopping on a lattice according to predefined rules, and interacting in particular through exclusion rules.

Cellular automata are defined not only by the specification of the underlying lattice, and of the hopping rules, but also by the update order, \textit{i.e.} the order in which the rules will be applied to the set of particles. It is known that update schemes can have a strong influence on the global dynamics of the system~\cite{rajewsky1998}. While random sequential update~\footnote{In the random sequential update, one particle at a time step is chosen randomly and updated.} is preferentially used in fundamental studies on out-of-equilibrium systems because of its closeness to continuous time dynamics, other updates with lower fluctuations are used in traffic applications. The most widely used is the \textit{parallel} update~\cite{chowdhury_s_s2000}, for which all particles are updated at the same time. In contrast with the random sequential update,  the time step of parallel update can be given a physical meaning and be interpreted as a reaction time.

Some other update schemes have been proposed,  such as the  random shuffle update~\cite{kessel2002} and the  frozen shuffle update~\cite{appert-rolland_c_h2011a}, both inspired by pedestrian applications. We shall emphasize that all these shuffle updates can be defined in terms of variables associated to the particles, and called \textit{phases}. The phases determine the order in which particles are updated. Both the \textit{random} and \textit{frozen} shuffle updates were investigated in detail in one dimension~\cite{wolki_s_s2006,wolki_s_s2007,smith_w2007a,smith_w2007b,appert-rolland_c_h2011a,appert-rolland_c_h2011b,cividini_a_h2014b}.

In this paper, we want to explore the properties of these updates when applied to a two-dimensional model. As a test case, we chose to consider the evacuation of a square room. Pedestrians are modeled by a simple so-called \textit{floor field model}~\cite{burstedde2001b} that we will describe more precisely in section~\ref{section:model}.

When the density increases, a crossover occurs between  low- and  high-density regimes. We will characterize this crossover, and the outflow in both regimes, including its dependency on the strength of the floor field. We will also discuss briefly the shape of the queue in front of the exit.

Another aim of this paper is to illustrate by an example how it is possible to introduce a dynamics  for the phases defining the update, and how this can modify the properties of the model. Eventually we shall discuss these results in view of pedestrian applications.

This article is organized as follows.  In section~\ref{section:model}, we define the model in detail, and present the various shuffle update schemes used in the paper. In particular, we introduce some phase dynamics to define  a \textit{hybrid shuffle update}. Some theoretical and numerical results are presented in sections~\ref{section:numsimu} and \ref{section:highd}, where we study two different, low- and high-density regimes. In section~\ref{section:pedevac}, we turn to pedestrian applications and summarize some experimental results as well as modeling approaches from the literature. We discuss the most prominent features of evacuation flows that one would like to reproduce in models. In section~\ref{section:conclusion} we give the conclusion of this article. 

\section{Model}
\label{section:model}

In this article  we consider one of the simplest situations of pedestrians evacuating a room  in the tradition of cellular automata. The room is divided into cells.  Pedestrians are simply represented by `particles' hopping stochastically  from one cell to one of the neighbouring cells. Each cell can be occupied by at most one particle (`simple exclusion').  Several evacuation models of this kind have been proposed~\cite{burstedde2001b, kirchner2003a, kirchner_n_s2003, yanagisawa2009, schadschneider_s2011}, in which the particles are usually updated in parallel, an update also traditionally used in car traffic problems.

\subsection{Lattice geometry}

\begin{figure}
\begin{center}
  \includegraphics[width=0.6\textwidth]{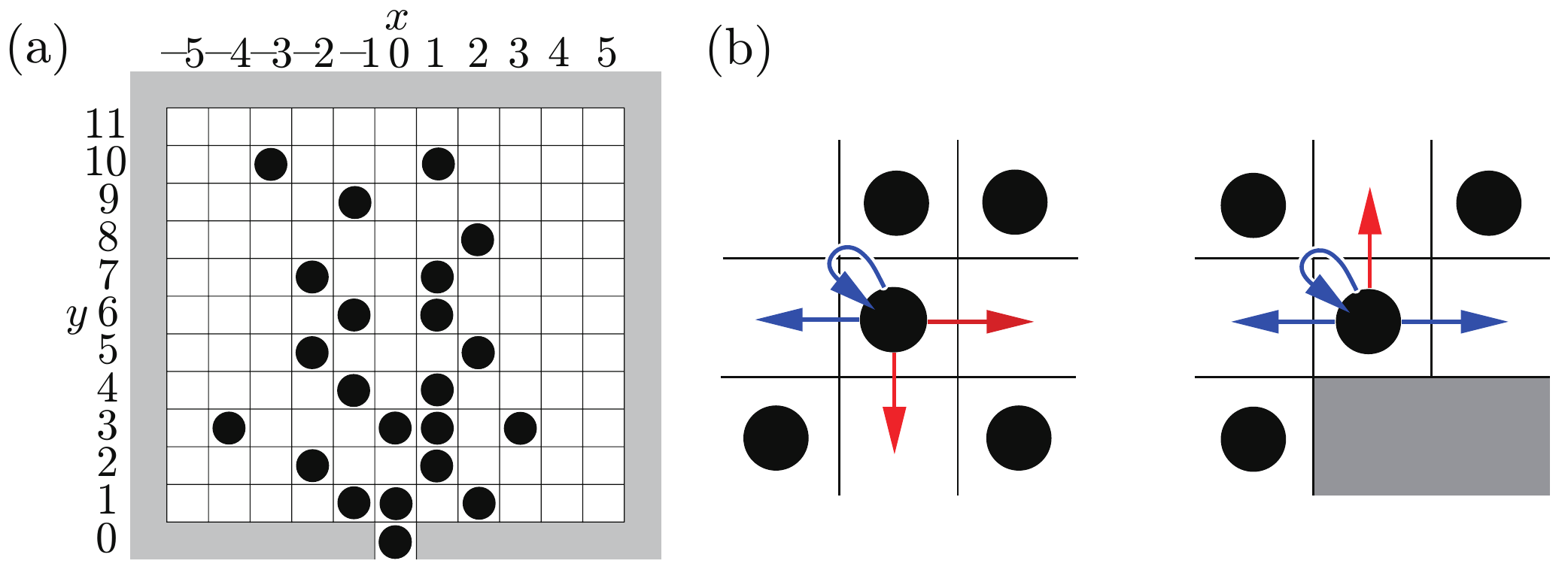} 
\end{center}
\caption{\small (a) Geometry of the room for $L = 11$, $l=5$  and (b) possible hopping directions in two particular configurations. The dots  represent particles and the grey cells are walls.  The cells of the room have coordinates $(x,y)$ with $|x| \leq l$ and $0 < y \leq 2l+1$,  and  $(0,0)$ is the exit.  The arrows represent the allowed hops for the particles in the centers.  In the case of hybrid shuffle update which is defined in subsection~\ref{subsection:hybrid}, the phase $\tau$  stays the same for the hops  with blue arrows and is redrawn in the case of red arrows.}  
\label{fig:room}
\end{figure}

We consider a square room of $L \times L$ cells, where $L =2 l + 1$ is an odd number, see figure~\ref{fig:room} for $L=11$. The room has a unique exit located in the middle of a side, whose width is just the size of one cell.  The cells of the room will be designated by their coordinates $(x,y)$ with $x = -l,-l+1,\ldots,l$ and $y = 1,\ldots,2l+1$, with a supplementary cell $(0,0)$ representing the exit.

In the following, particles will be allowed to hop from one cell to its empty neighbours, and a hop from $(x,y)$ to $(x',y')$ will sometimes be denoted as $(x,y) \rightarrow (x',y')$. This latter notation implicitly assumes that $x$, $y$, $x'$ and $y'$ are such that the hop is indeed allowed, \textit{i.e.} $(x,y)$ and $(x',y')$ are neighbours, $(x,y)$ is occupied and $(x',y')$ is empty.

\subsection{Hopping probabilities}
\label{subsection:partmot}

At initial time and during the whole time evolution the  simple exclusion constraint is verified, \textit{i.e.} there can be   at most one particle on each cell.

We direct the particles    towards the exit by  using  a static floor field~\cite{burstedde2001b}: The probability that each particle hops to a certain target cell is determined by the distance between the target cell and the exit of the room. We define a parameter $k \geq 0$ that quantifies the strength of the   attraction towards the exit.
We denote the Euclidean\footnote{
See~\cite{nishinari2010} for a comparison of the queue shapes when Euclidian or Manhattan distances are used.} norm by $|\mathbf{r}| \equiv \sqrt{x^2+y^2}$ for   $\mathbf{r} = (x,y)$. To each cell of the room we associate a weight 
\begin{equation}
\label{eq:defw}
 w(\mathbf{r}) \equiv \mathrm{e}^{-k |\mathbf{r}|}  .
\end{equation}
A particle on cell~$(x,y)$ chooses its target cell   among all of its von Neumann neighbourhood (cells~$(x \pm 1,y)$ and~$(x,y \pm 1)$) that are empty and the cell it already occupies.  Among the possible targets, site $\mathbf{r}'$ is chosen with probability 
\begin{equation}
\label{eq:defprobahop}
 p(\mathbf{r}') = Z^{-1} w(\mathbf{r}')   
\end{equation}
with the normalization $Z = w(\mathbf{r}) + \sum_{\mathbf{r}'} w(\mathbf{r}')$, where the summation   runs over all the empty von Neumann neighbours of $\mathbf{r}$. Figure~\ref{fig:room} (b)  summarizes the possible   hops of a particle.

Finally, cell~$(0,0)$ is an exception to the rule~(\ref{eq:defprobahop}): a particle standing on~$(0,0)$ exits the system with probability $1$. In a larger simulation, its hopping probability would be determined by the downstream configuration.

\subsection{Shuffle update schemes}
\label{subsection:updates}

When we use cellular automata, we need to specify an update scheme to fully define the model.   In this paper, we shall exclusively consider shuffle updates.

Shuffle updates are sequential updates, \textit{i.e.} particles are updated one after the other. Here we chose to formulate these updates by associating to each particle $i$ a phase  $ \tau_i\  ( 0 \le \tau_i < 1  ) $.  Then the initialization procedure must not only define the position of particles at initial time, but also associate to each particle a phase drawn from a uniform distribution. If the system is open, injected particles are given a phase when they enter the system.

By definition of the shuffle updates, at each time step the particles are updated in order of increasing phases:
\begin{equation}  \tau_{i_1} <\tau_{i_2} < \cdots < \tau_{i_N} , \end{equation}
where $N$ is the number of particles present at a given time in the system. Note that with shuffle updates, if successive sites are occupied by particles with increasing phases, the set of particles can move as a whole one step forward within one single time step.

Equivalently  to the previous definition, in a continuous time picture, one can consider that each particle $i$ is updated at  time $s+\tau_i$, where $s$ is the discrete time. 

Two variants of shuffle updates have been considered in the literature. In the  frozen shuffle update, phases are kept unchanged during the whole simulation. By contrast, in the  random shuffle update, phases are drawn anew at the beginning of each time step.

Shuffle updates have low fluctuations, in the sense that each particle is updated exactly once per time step. Still, for a given particle, the time between two updates can fluctuate between $0$ and $2$ for the  random shuffle update, while it is always exactly one in the frozen shuffle update. Indeed, in the  frozen shuffle update, a particle $i$ is updated at times $\tau_i ,\tau_i +1,\tau_i +2,\ldots$.
 
When particles hop on a one-dimensional lattice with a constant hopping probability $p$, the model  is equivalent to the so-called totally asymmetric simple exclusion process~\cite{chou_m_z2011}. In this  1D  case, the fundamental diagram, \textit{i.e.} the current  $J$ as a function of $\rho$, has already been calculated analytically for both updates in the deterministic case ($p=1$). For the  random shuffle update~\cite{wolki_s_s2006,smith_w2007a}, the  current with deterministic hopping    reads
\begin{equation}
J(\rho,p=1) = 
\left\{ \begin{array}{ll}
\rho  & \textrm{for} \; \rho \le 1/2,  \\
\frac{\rho(1-\rho)}{2\rho-1}\left[\exp\left(\frac{2\rho-1}{\rho}\right)-1\right]
\quad \quad  & \textrm{for} \; \rho > 1/2  . 
\end{array}
\right.
\end{equation}
 In particular,  this current has a maximum $ J^{\rm 1D}_{\rm R} =  \frac{1}{2}$ at $ \rho  = \frac{1}{2}$.  For the  frozen shuffle update~\cite{appert-rolland_c_h2011a,appert-rolland_c_h2011b}, the fundamental diagram reads 
\begin{equation}
J(\rho,p=1) =
 \left\{ \begin{array}{ll}
    \rho  & \textrm{for} \; \rho \le 2/3 , \\
   2(1-\rho) \quad \quad  & \textrm{for} \; \rho > 2/3
\end{array}
\right.
\end{equation}
with a maximum $J^\mathrm{1D}_{\rm F} = 2/3$.  This maximum corresponds to configurations where, after a short transient, particles form \textit{platoons}. More precisely, an observer staying on  a site and recording its occupation and the phases $\{\tau_i \}$ of the particles that occupy it at each time step would see a sequence of particles with increasing phases, then a hole, then another sequence of particles with increasing phases, and so on. A sequence of particles without a hole between them will be identified as a platoon.  If we denote the average number of particles in a platoon by $\nu$, then $\nu$ particles will exit every $\nu +1$ time steps,  thus giving a maximal current  $\frac{\nu}{\nu+1}$.  For phases drawn from a uniform distribution of values between 0 and 1, $\nu$ can be shown to be equal to $2$~\cite{appert-rolland_c_h2011a}, hence the maximal current value.

We see that $J^\mathrm{1D}_{\rm F} > J^\mathrm{1D}_{\rm R}$.
Indeed, the repeated changes in the updating order of  random shuffle update   provides a supplementary source of blockings between particles, leading to less regular motion.

For the  frozen shuffle update,  one must keep in mind that the flow may depend on the exact realization of the updating order, \textit{i.e.} on the specific choice for the phases. This is the case for closed systems, or for the evacuation problem of this paper, or more generally for any system that involves a bounded number of particles whatever the simulation duration is. The phases are like a \textit{frozen disorder}, and one needs to  take  ensemble averages to get the mean behavior~\cite{appert-rolland_c_h2011a}.

By contrast, in open systems with incoming flows~\cite{appert-rolland_c_h2011b}, the time average would coincide with an ensemble average over the  frozen disorder, \textit{i.e.} over the phases, and a single simulation would be enough.

\subsection{The hybrid shuffle update}
\label{subsection:hybrid}

Systems with  random shuffle  and frozen shuffle updates  do not behave in the same way. For the first one, the renewing of the updating order introduces some randomness in the bulk particle dynamics while for the second, the disorder is injected at the boundaries and the bulk dynamics is deterministic (if the hopping rules are deterministic).

It is also possible to combine both aspects by introducing a dynamics of the phases themselves. The choice of this dynamics is not unique. We will present one possible choice of hybrid shuffle updates in this section\footnote{In reference~\,\cite{arita_c_a2014} we have introduced another hybrid shuffle update. The main difference is that here target sites are necessarily empty, hence the phase dynamics cannot be based on the occupancy of the target site as in reference~\,\cite{arita_c_a2014}. The choice made in this article leads to higher flows in high density regime.}. This choice was guided by the request to have an update behaving similarly to the  frozen shuffle update  when the particle density is low, and closer to the  random shuffle update  inside congestions. We will discuss further our choice in section~\ref{section:pedevac}, but first we will define the hybrid shuffle update of this paper.

We introduce the following rule. If a particle $i $ attempts to hop from $(x,y)$ to $(x+a,y)$ (or $(x,y+a)$),  and if the neighbours of the arrival cell $(x+a,y\pm1)$ (or $(x \pm 1, y+a)$) are both occupied by other particles,   then the phase of the hopping particle $\tau_i $ is redrawn
from a uniform distribution on $ 0 \le \tau_i <1   $.

The phase dynamics has no incidence on the motions of particles, \textit{i.e.}  the hops are  always allowed when particles chose a neighbouring empty cell,  according to the probability~(\ref{eq:defprobahop}). Only the order of update will be modified, starting from the next time step.

As particles on $(0,0)$ are supposed here to quickly step out, and thus not to be in the congestion anymore, we do not redraw the phase when a particle hops to $(1,0)$ even if $(0,0)$ and $(2,0)$ are occupied.

Note that, in the one dimensional case, no renewal of phases occurs by definition.  Therefore the hybrid shuffle update scheme is equivalent to the frozen one in one dimension, and we have  
\begin{equation}\label{eq:H=F-in-1d} J^\mathrm{1D}_\mathrm{H} = J^\mathrm{1D}_\mathrm{F} = \frac{2}{3} . \end{equation}

As we mentioned, the shuffle update schemes presented in section~\ref{subsection:updates} were already studied in detail in one dimensional systems (at least for the deterministic TASEP). Therefore one of our  particular interests in this article is to investigate their behavior in a typical two dimensional system, \textit{i.e.} an evacuation from a room though a small exit.  In the next two sections, we shall explore the behavior of the evacuation model, and distinguish
 two dynamical regimes.
 
\section{Low-density regime }
\label{section:numsimu}

 At initial time $t=0$, $N$ particles are dropped randomly  on the lattice of figure~\ref{fig:room}  and their phases $\{\tau_i \}$ are drawn randomly from a uniform distribution.  In this section and the next one, we show simulation results that we have performed  with the room size $L^2=51^2$ and various initial numbers $N$ of particles, using either the random, frozen or hybrid shuffle update scheme.  We will be interested in the total evacuation times $T$, \textit{i.e.} the time when the last particle leaves the room,    and the flux $J$ of particles   exiting the room, which is identical to the current between $(0,1)$ and $(0,0)$.

\begin{figure}
\begin{center}
  \includegraphics[width=0.3\textwidth]{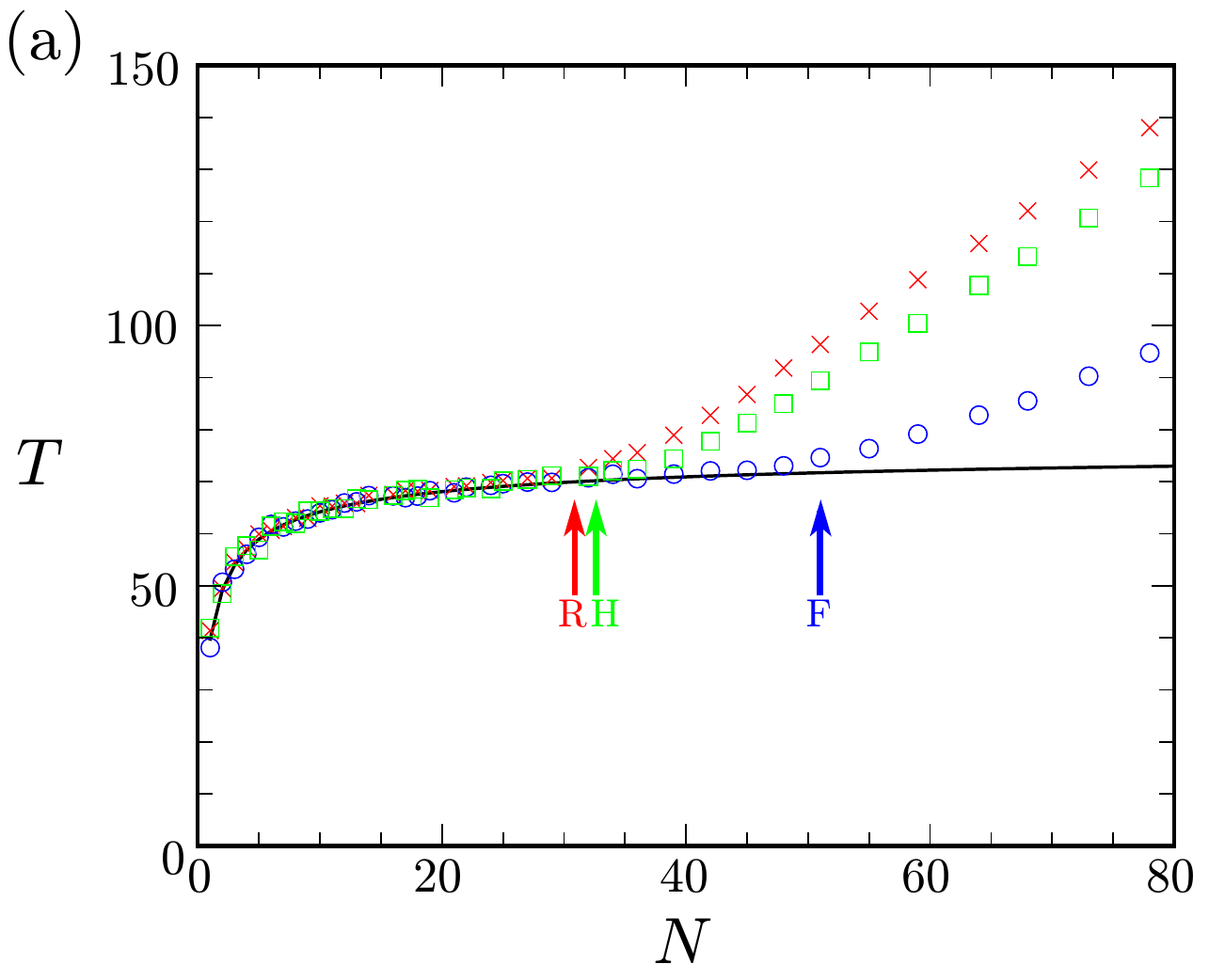} 
  \includegraphics[width=0.3\textwidth]{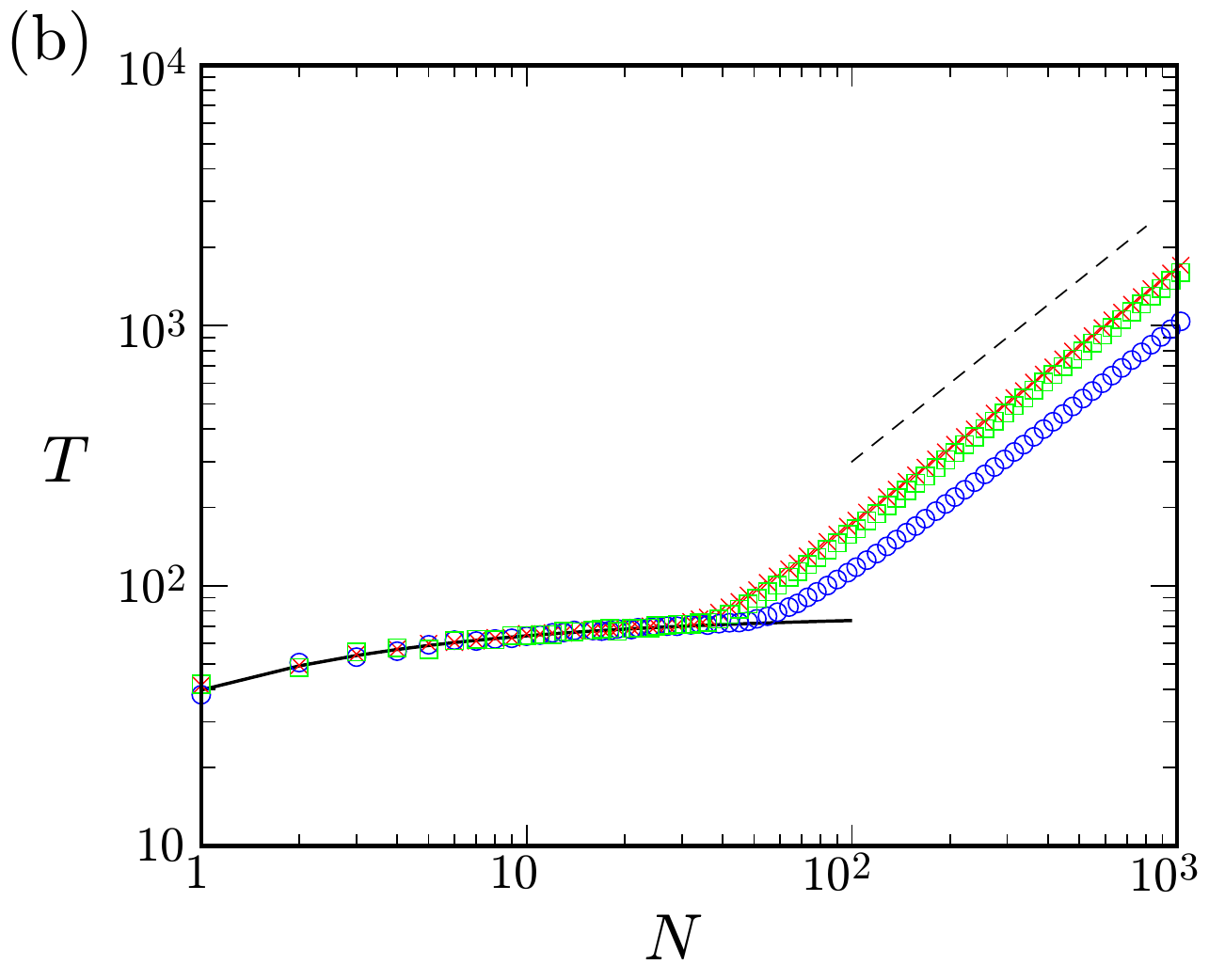}  \\ 
  \includegraphics[width=0.3\textwidth]{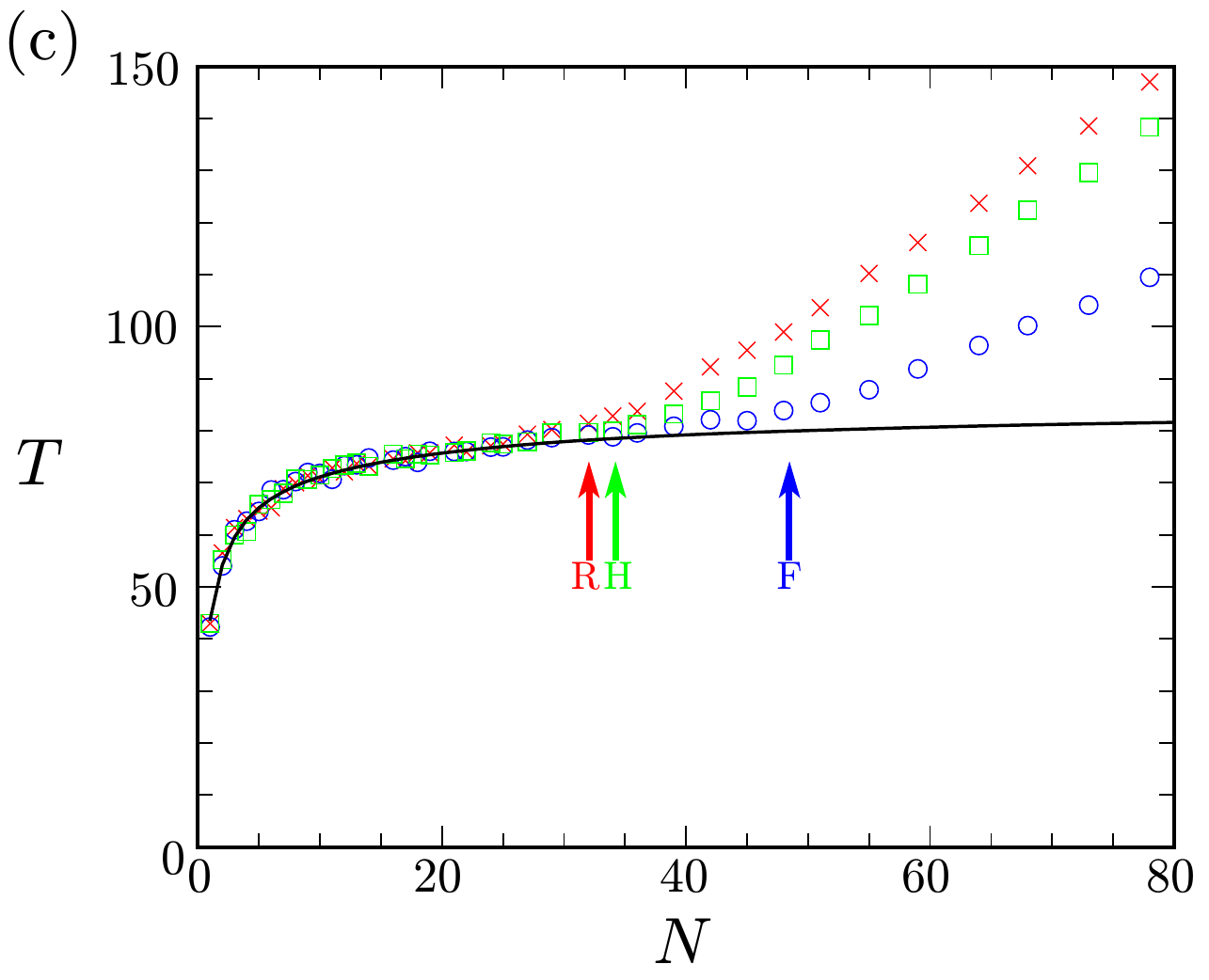} 
  \includegraphics[width=0.3\textwidth]{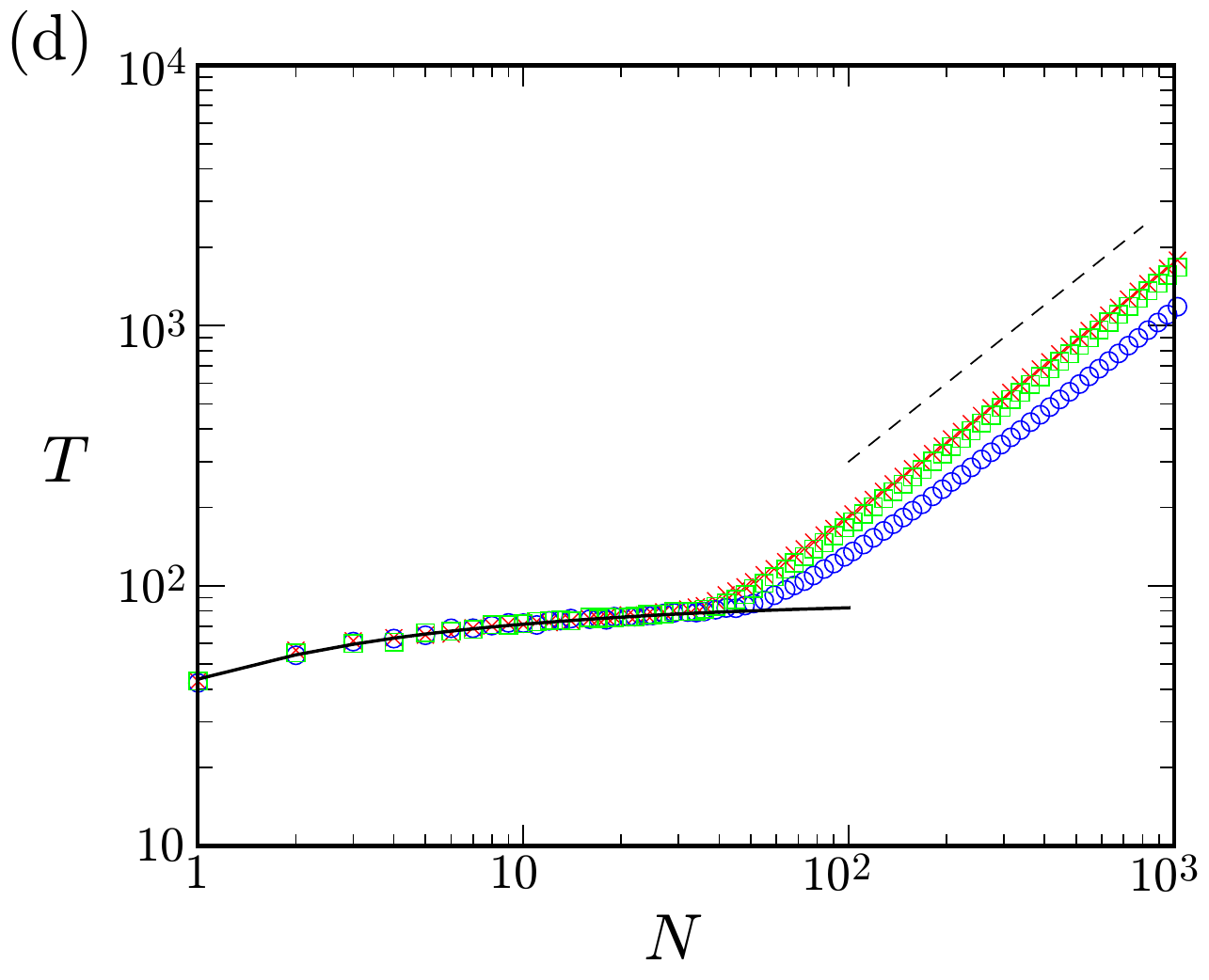} 
\end{center}
\caption{\small Total evacuation time $T$ versus the number of particles $N$ with random (red $\times$), frozen   (blue $\bigcirc$)  and hybrid  (green $\square$)   shuffle updates in linear (left) and logarithmic (right) scales. The room size is $L^2=51^2$, and the parameter $k$ is chosen as  $ k=\infty $ (a,b)  and $k=3$ (c,d). 
The plot markers were obtained by averaging over $10^2$ simulation runs.
  The black solid lines in (a,b) and (c,d)
 correspond to (\ref{eq:totalevactimeld})
and  (\ref{eq:EmaxTTT}), respectively. 
 For comparison, we drew dashed lines $ J= \mathrm{const.}\times t $ in (b) and (d). 
In (a) and (c), the arrows indicate  the crossover points (\ref{eq:crossoverN}) between 
the low- and high-density phases,
see  the last paragraph of Subsection~\ref{subsection:maxcurrent}.  
}  
\label{fig:totalevactime}
\end{figure}

Figure~\ref{fig:totalevactime} shows the total evacuation time $T$  versus  the initial number $N$ of particles. When the number of particles increases, there is a  smooth  transition between two qualitatively different regimes, referred to as the \textit{low-density} and \textit{high-density} regimes.  Let us first study the low-density regime.

For low   densities, the evacuation time is observed to be independent of  the update scheme. Indeed, in this free flow regime, particles  exit almost freely, and the velocity of non-interacting particles  is the same for the three updates.

The increase of the total evacuation time  $T$ with $N$ is then simply due to the increase of the average of the distance between the farthest particle and the door. A particle starting on cell $(x,y)$ is at a Manhattan distance $d=|x|+y$ from the exit. Neglecting all interactions with other particles, and assuming $k = \infty$, this particle will need a time $t=d+1=|x|+y+1$ to evacuate the room. Recalling  that  the shape of the room is square with $L^2$ cells ($L=2l+1$) and  that  the exit is located at  the  middle of one side,   the number $n(d)$ of cells located at a  Manhattan distance $d$ from the exit is found to be
\begin{equation}
 \label{eq:exprn}
  n(d) = \left\{  \begin{array}{lll}
               2 d -1 \qquad    & \mbox{ for }  &1 \leq d \leq l,  \\
               2l+1 \qquad     & \mbox{ for } & l+1 \leq d \leq 2l+1,  \\
               6l+4-2d \qquad  & \mbox{ for } &2l+2 \leq d \leq 3l+1  .
              \end{array} \right.  
\end{equation}
Note that the $n(d)$ takes a maximum value $L$ when $l+1 \leq d \leq 2l+1 $.
 Then we shall show that the average total evacuation time $T$ is given by
\begin{equation}
 \label{eq:totalevactimeld}
T = 1 +  \bigg(\!\!\!\begin{array}{c} L^2 \\ N  \end{array}\!\!\!\bigg)^{-1} \, 
\sum_{d = 1}^{3l+1} d
 \bigg[  \bigg(\!\!\!\begin{array}{c} \sum_{d'=1}^d n(d') \\ N  \end{array} \!\!\!\bigg)  -
  \bigg(\!\!\!\begin{array}{c} \sum_{d'=1}^{d-1} n(d') \\ N    \end{array} \!\!\!\bigg)   \bigg]  ,
\end{equation}
where $\bigg(\!\!\!\begin{array}{c}  a \\ b  \end{array}\!\!\!\bigg) = \frac{a!}{b!(b-a)!}$ is a binomial coefficient. In equation~(\ref{eq:totalevactimeld}) the difference of binomial coefficients is the number of initial configurations where the farthest particle is at distance exactly $d$ from the exit. One obtains the probability that the farthest distance is $d$, by dividing it by the total number of initial configurations $\bigg(\!\!\!\begin{array}{c} L^2 \\ N  \end{array}\!\!\!\bigg)$,  and then the summation over $ d=1,\dots, 3l+1 $ with multiplication by $d$ gives the average farthest distance.   
The expression (\ref{eq:totalevactimeld}) of $T$ versus  $N$ is plotted for $L=51$ in figure~\ref{fig:totalevactime} (a). We observe that the agreement is good when $N$ is small, \textit{i.e.} as long as particles (almost) do not interact. However, the true evacuation time becomes larger than  predicted by (\ref{eq:totalevactimeld})  as $N $ increases, as a result of the collective effects occurring in the high-density regime.

We now discuss in more detail the crossover towards this large $N$ regime. Because of the geometry of the room, the number of particles arriving at the exit is not constant in time. A jam will form at the exit if the incoming flow $ J^\mathrm{in}(t)$ becomes larger than the maximal current that the exit bottleneck can sustain. This limit value is called the capacity of the bottleneck, and is equal to the outflow in the high-density regime, denoted by $J^\mathrm{2D}$. The  value of $J^\mathrm{2D}$ will be determined in subsection~\ref{subsection:maxcurrent}. For now it is enough to know that $J^\mathrm{2D}$ takes some numerical value that depends on the update scheme, but not on $L$ or $t$.

Let us first compute the incoming flow $J^\mathrm{in}$ of particles. For the time being, we assume that particles move without interacting with each other. In the case $k=\infty$, the time $t$ needed by a non-interacting particle to arrive at the exit site $(0,0)$ is equal to the Manhattan distance $d=|x|+y$ between its initial position $(x,y)$ and the exit. Then, for an initial uniform density $N/ L^2$ of particles, the inflow coming to the exit at time $t$ is
\begin{equation}\label{eq:Jin}
   J^\mathrm{in}(t)  =  n(t) \times N/ L^2 , 
\end{equation}
where $n(t)$ is given by (\ref{eq:exprn}). Between an initial increase and a final decrease, the function $J^\mathrm{in}(t)$ reaches a plateau value $J^\mathrm{in}_{max} = N/L$.

When $ J^\mathrm{in}_{max} > J^\mathrm{2D} $, congestion occurs, \textit{i.e.} one enters the high-density regime.  
The crossover thus occurs, when the number of particles is 
\begin{equation}
\label{eq:crossoverN}
   N_c  = J^\mathrm{2D} L . 
\end{equation}
In the next section we shall focus on the high-density regime 
and determine the update-dependent value of $J^\mathrm{2D}$.
This will allow us in particular to locate the crossover between
the two regimes.

For  finite $k$,  a similar transition is observed, see figure~\ref{fig:totalevactime} (c,d).  The form (\ref{eq:totalevactimeld}) is no longer valid in the low-density regime for finite $k$.  Instead of this we compared the genuine evacuation times of $N$  particles (colored markers) with the single-particle problem (\textit{i.e.} initially only one  particle is put   randomly in the room). The solid lines were  obtained by  averaging the maximum of   evacuation times $T_i$ of  $N$ independent simulation runs of the single-particle problem, \textit{i.e.} formally 
\begin{equation}\label{eq:EmaxTTT}
     \mathbb E [\max\{T_1, \cdots, T_N \}] . 
\end{equation}

\section{High-density regime }
\label{section:highd}

In this section we study the evacuation from  the room in the high-density regime.

\subsection{Outgoing currents ($ k=\infty $)}
\label{subsection:maxcurrent}

\begin{figure}
\begin{center}
    \includegraphics[width=0.3\textwidth]{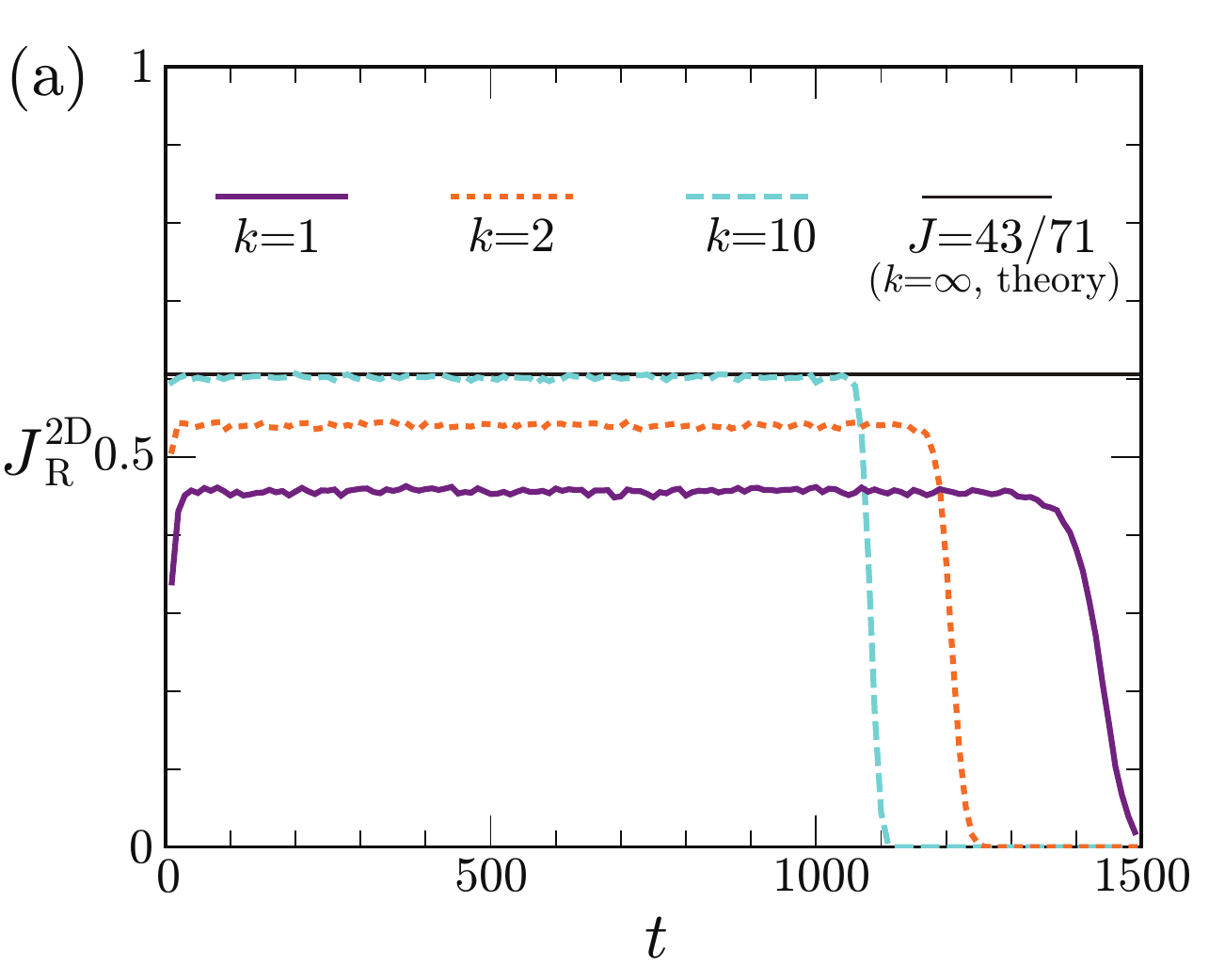} 
     \includegraphics[width=0.3\textwidth]{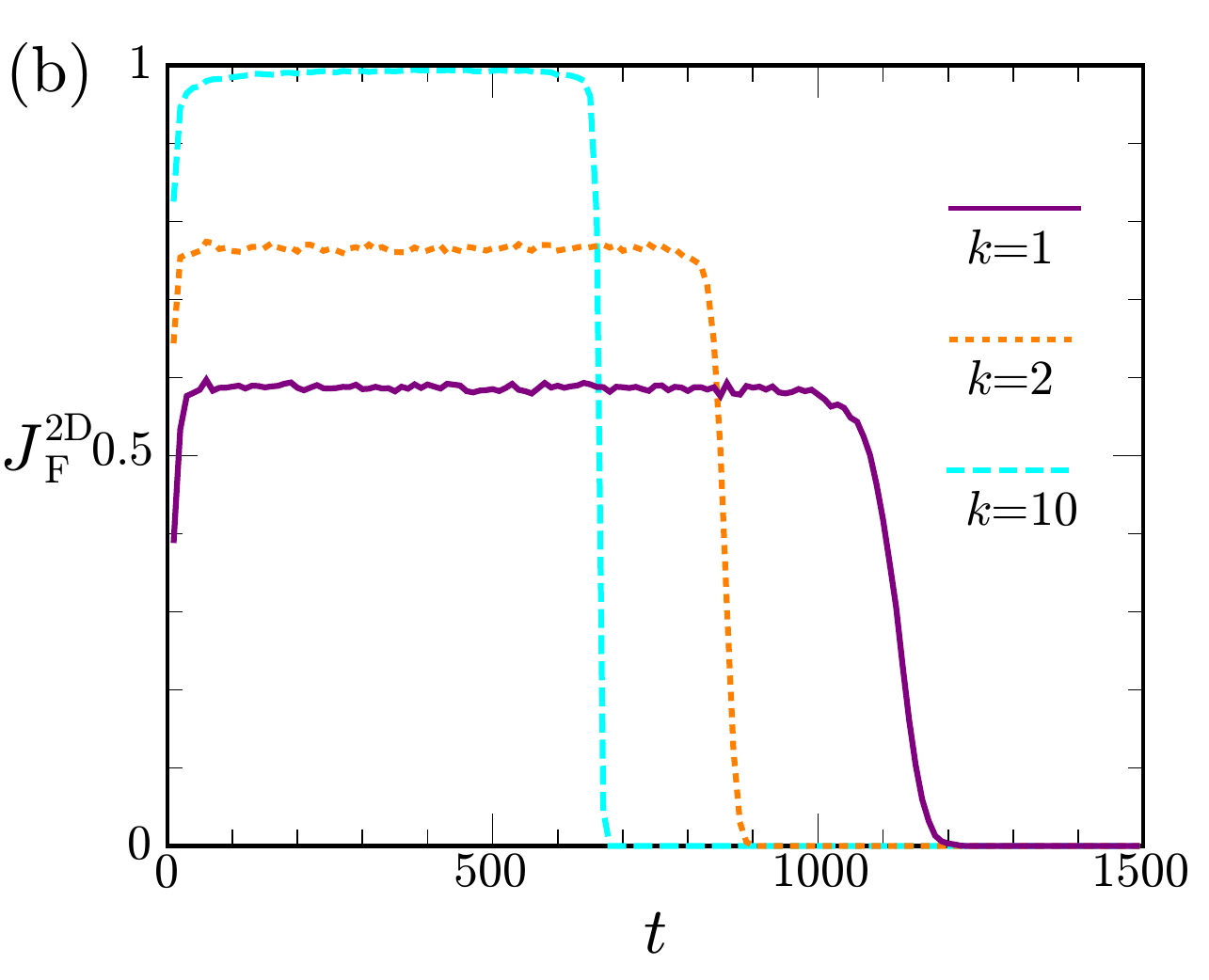} 
     \includegraphics[width=0.3\textwidth]{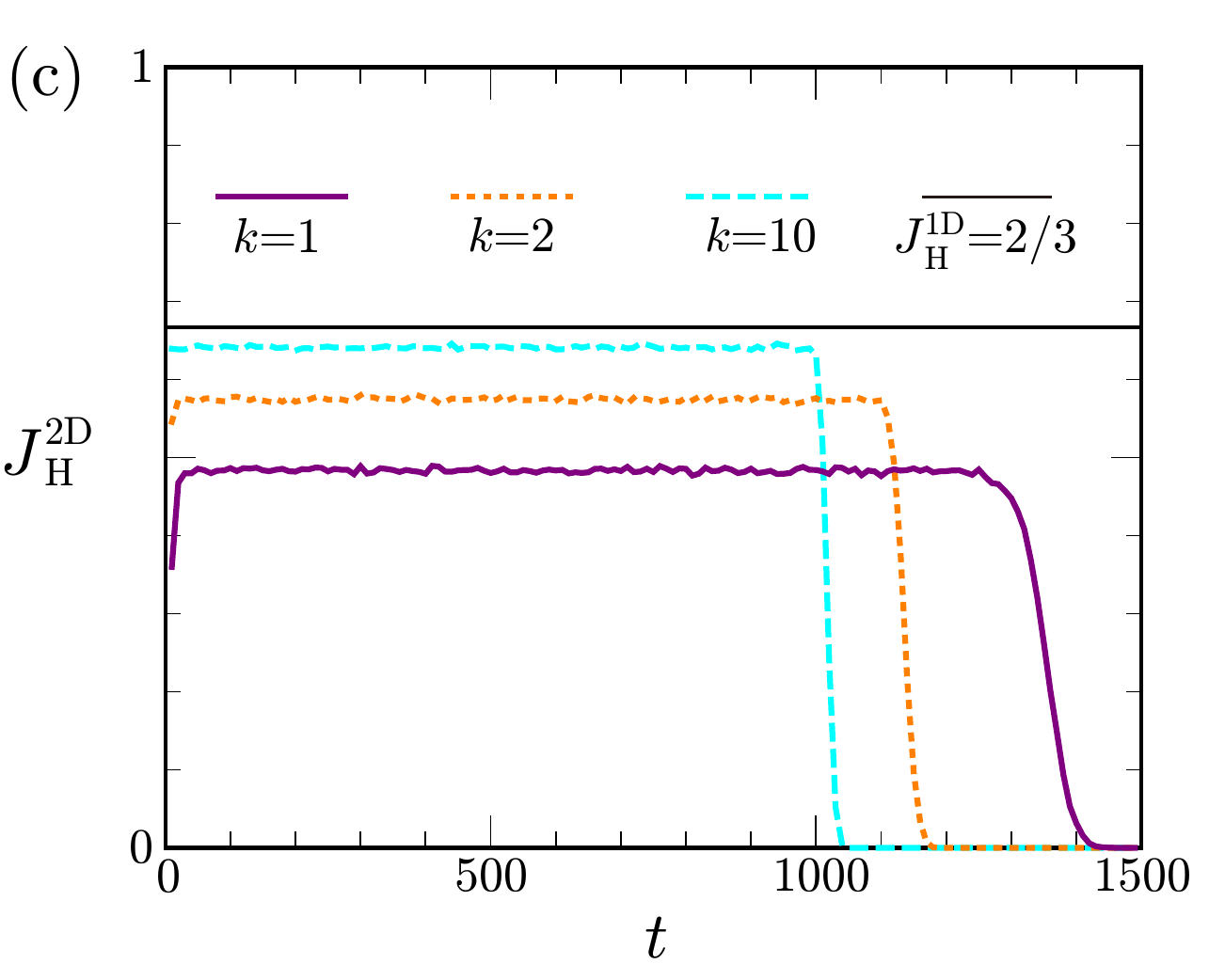} 
     \end{center}
\caption{Outflows $ J^\mathrm{2D}_X  $
as  functions of time for random shuffle (a), frozen shuffle (b) and hybrid shuffle (c)
updates  for three values of $k = 1$, $2$ and $10$ and for a room of size
$L^2=51^2$ with quarter-filling, $N=(L^2-1)/4=650$   (the side $L$ always has an odd number of sites). We are thus clearly in the high-density regime. The theoretical predictions for $k=\infty$ are   $J^\mathrm{2D}_\mathrm{R} = 43/71$ for random shuffle update and $J^\mathrm{2D}_\mathrm{F} = 1$ for frozen shuffle. For comparison, we also drew the line $J^\mathrm{1D}_\mathrm{H}=2/3$ for the hybrid case.  
\label{fig:currentt}
}
\end{figure}

From figure~\ref{fig:totalevactime} it can be seen that in the high-density regime the evacuation time grows linearly with the number of particles. Moreover, it can be checked in figure~\ref{fig:currentt}
 that the particle current flowing out of the system (the outflow $J^\mathrm{2D}_X$)
  is constant in time except in the initial and final transients, so that the evacuation time is approximately given by
$T=N/J^\mathrm{2D}_X$, where $J^\mathrm{2D}_X$ depends on the update schemes, 
\textit{i.e.} 
$X=$R (random shuffle),  F (random shuffle) and H (hybrid shuffle).
It remains to compute the value of this current for these three updates,
 which is done for $k = \infty$ in the following paragraphs.

We start with the case of random shuffle update. The outflow will be determined by the dynamics around the exit.
We therefore write an approximate master equation for the $P^a_{\,  b} (t)$, where the letters $a$ and $b$ stand for the occupations of cells $(0,1)$ and $(0,0)$, respectively,  \textit{e.g.}
we write $a=1 $ or $a= 0$ depending on whether $(0,1)$ is occupied or unoccupied. 
Since we expect that there is a queue around the exit, we suppose that    $(1,1)$, $(0,2)$ and $(-1,1)$ are always occupied.  Under this first order approximation and 
with the simple choice $k = \infty$, cells  $(0,1)$ and $(0,0)$ are never  simultaneously  empty at integer times and we can write a closed equation for the three remaining probabilities as 
\begin{equation}
 \label{eq:meqrsu}
\vec P (t+1) = M \vec P (t), \ {\rm where} \ 
\vec P (t)  = \Bigg( \begin{array}{l} P^1_{\, 0}(t) \\ P^0_{\, 1}(t) \\ P^1_{\, 1}(t) \end{array} \Bigg) ,   \ 
M =
\Bigg( \begin{array}{ccc} 0 & 1 & 1/2 \\ 1/4 & 0 & 1/5 \\ 3/4 & 0 & 3/10\end{array} \Bigg) .
\end{equation}
Let us explain for instance how the first column of the matrix can be obtained. It corresponds to the transition probabilities from state $_0^1$ to  states $_0^1$,  $_1^0$ and  $_1^1$. Starting from state $_0^1$, the particle in $(0,1)$ will surely hop to $(0,0)$ during the time step.  Consider the particles occupying sites $(0,1)$, $(1,1)$, $(0,2)$ and $(-1,1)$. If the particle on $(0,1)$ draws the highest phase (probability $1/4$), the three other particles will be blocked and the state at the next timestep will be $_1^0$ whereas in the opposite case (probability $3/4$) at least one particle from $(1,1)$, $(0,2)$ or $(-1,1)$ will be able to hop towards $(0,1)$ and the next state will be $_1^1$.

By calculating the stationary solution   
\textit{i.e.}  $\vec P = M \vec P $, we find 
a stationary current 
\begin{equation}\label{eq:JR=0606}
J^\mathrm{2D}_\mathrm{R}=  P^0_{\, 1}+ P^1_{\, 1}
 = \frac{43}{71}    =0.60563\dots . 
 \end{equation} 
This prediction is in good agreement with the simulations,
 see  figure~\ref{fig:currentt}(a).

Let us turn to the frozen shuffle update. 
For the 2D case,  platoons  are also formed similarly to the 1D case: 
An observer staying on the exit site and recording its occupation and the phases $\{\tau_i \}$ of the particles that occupy it at each time step would see a sequence of particles with increasing phases, then a hole, then another
sequence of particles with increasing phases, and so on. A sequence of particles without a hole between them will be identified as a platoon. 
Platoons are self-organized in a stronger sense than the 1D case. 
The ordering  of the particles is not conserved  in 2D.
 And thus  they  become very large.  Numerically we observe that 
the average  size of these structures keeps growing with the number of particles in the room. 
 This fact   leads  to the current  
\begin{equation} \label{eq:JF=1}
   J^\mathrm{2D}_{\rm F} = 1
\end{equation}
which is expected to become the exact  value  in the limit $ N\to \infty $.

In the high-density regime of the two-dimensional evacuation, the hybrid shuffle update is very similar to the random shuffle update because there is a queue near the exit. When a particle $i$ hops, most of the neighbouring cells are   
occupied, so that its phase $\tau_i$ will  often be redrawn. The outflow  of the hybrid case  is therefore similar to that of the random case, as shown in figure~\ref{fig:currentt}. The remaining difference comes from the fact that,
in the hybrid case, phases are often but not systematically redrawn. Still, as we shall see below, it is enough to strongly limit the 2D outflow as we wanted. While we were able to predict the flow precisely for the random case in~(\ref{eq:meqrsu}), it is not so easy to do it for the hybrid shuffle update, because of the remaining `correlations' of the phases. By simulation results, the two dimensional current for the hybrid case is measured as 
\begin{equation}\label{eq:JH=064}
  J^\mathrm{2D}_\mathrm{H} \approx 0.64 . 
\end{equation}
The maximal one dimensional flow $J^\mathrm{1D}_\mathrm{H}$ for the hybrid shuffle update scheme is equal to $2/3$, as stated in (\ref{eq:H=F-in-1d}). The interactions between particles result  from  the dynamics of the phases $\{\tau_i \}$.

Now we go back to the problem of the crossover between low- and high-density regimes.  By substituting the values of the outflows in the  high-density regime (\ref{eq:JR=0606}), (\ref{eq:JF=1}) and (\ref{eq:JH=064})  into (\ref{eq:crossoverN}), we estimate the transitions between the low- and high-density regimes to occur at  $N_c \approx 30.8, 51$  and $32.6$   for the random,  frozen and  hybrid shuffle update schemes, respectively, for $L=51$.  These values agree with the numerical data, see figure~\ref{fig:totalevactime} (a).

\subsection{Dependency on $k$}

In the preceding subsection we studied the case $k = \infty$. Here we consider the effect of varying $k$ on the total evacuation time $T$.

\begin{figure}
\begin{center}
 \includegraphics[width=0.3\textwidth]{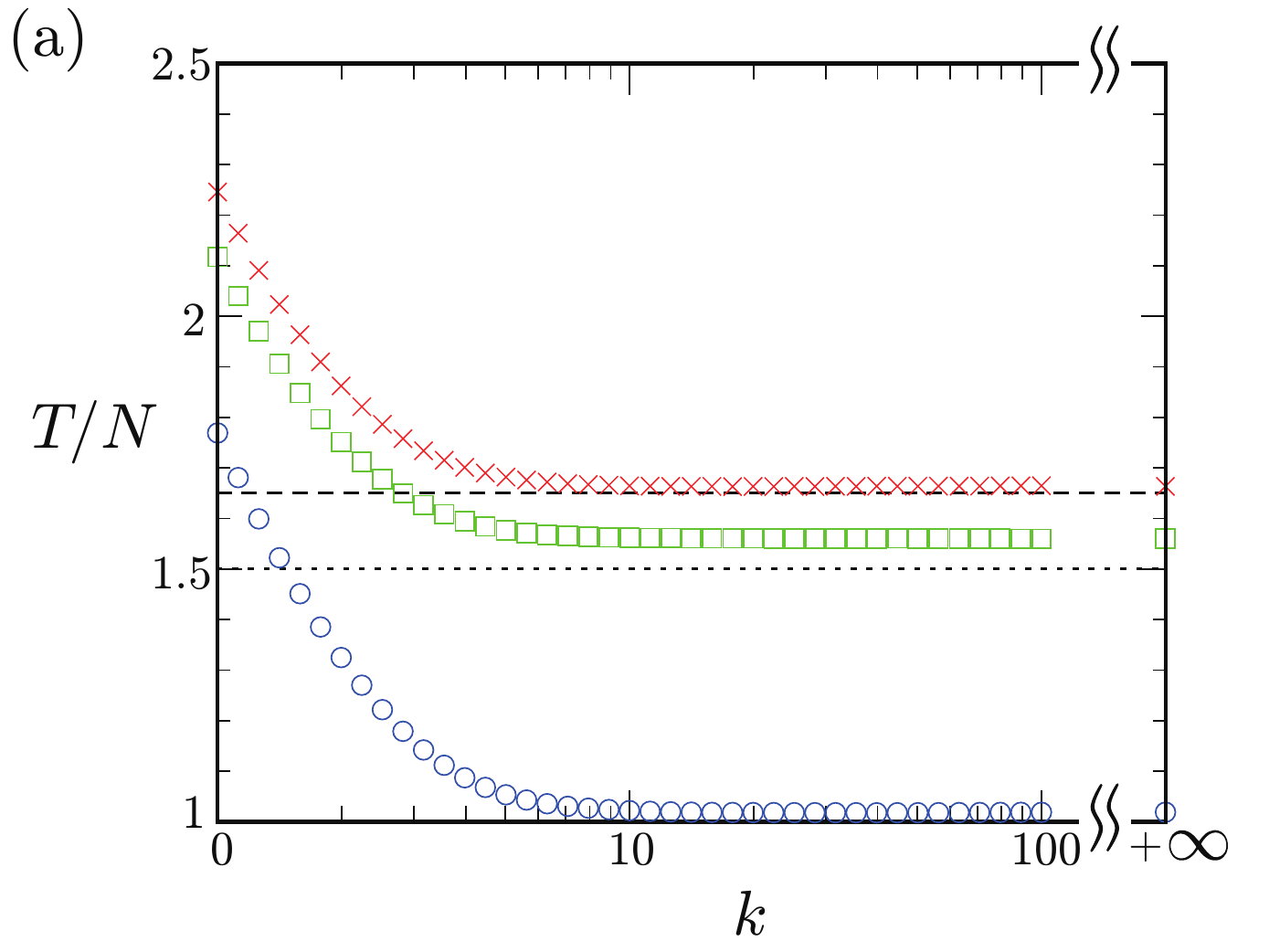}  
 \includegraphics[width=0.3\textwidth]{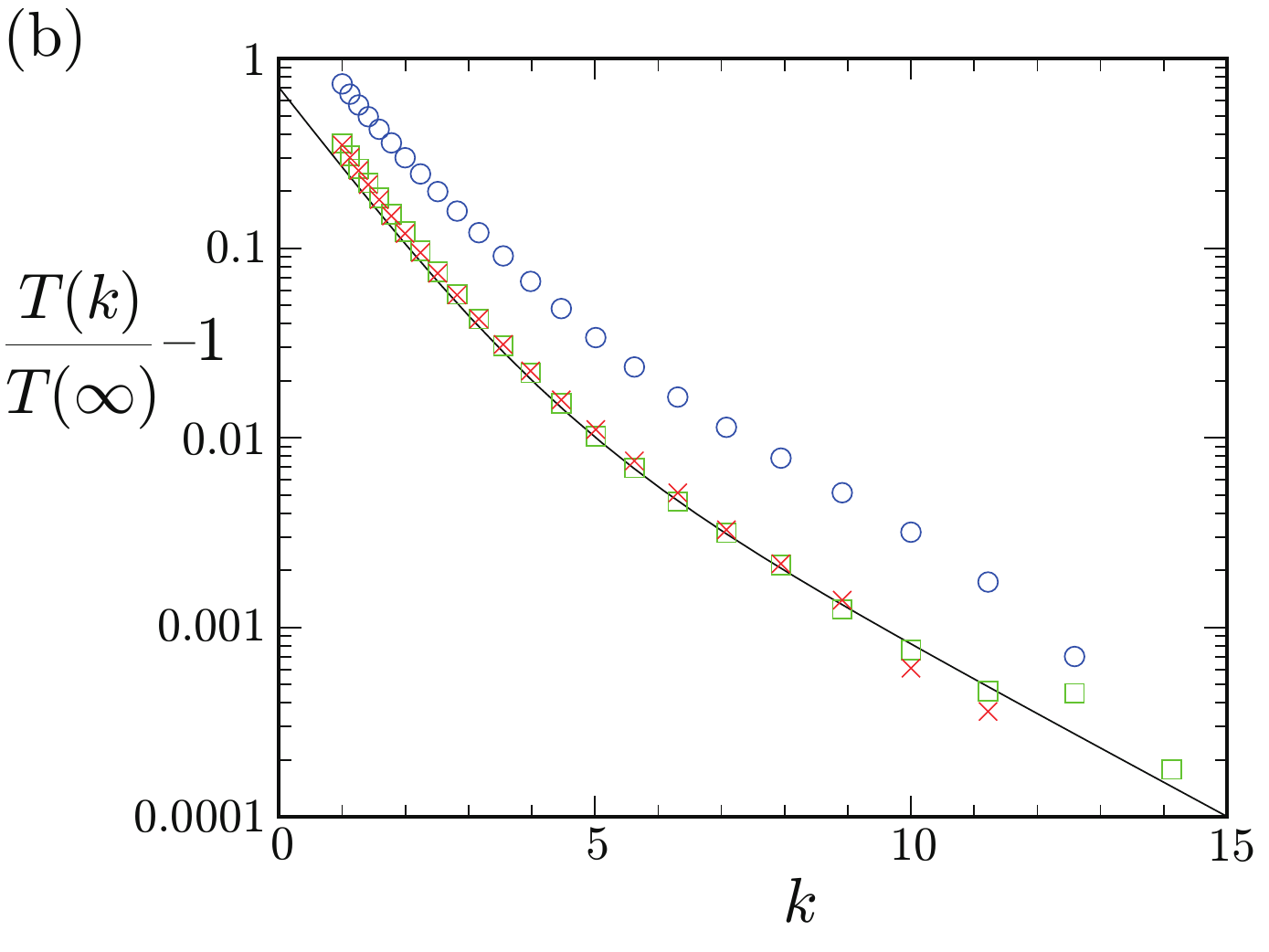}  
\end{center}
\caption{\small (a) Total evacuation time per particle as a function of $k$ 
and (b) its relative variation with $k$ 
  for the random (red $\times$),    frozen   (blue $\bigcirc$) and hybrid  (green $\square$) shuffle update schemes  with $L=51^2$ and $N=(L^2-1)/4$.
 For better visibility, we used logarithmic scales for $k$ and  $ T(k)/T(\infty) -1   $   in (a) and (b), respectively.
In (a) we observe the agreement with the predicted evacuation time $T/N=1 /J$ with equation (\ref{eq:JR=0606}) in the random shuffle case (dashed line).
For comparison,  we also drew the dotted line   $ T/N=1.5$, which is 
  obtained if  we assume that the outflow is identical to the 1D flow in the hybrid shuffle case.
To obtain the data points we averaged over $10^4$ simulation runs. 
Due to the finiteness of simulation runs,  fluctuations for large $k$ 
are observed in the log scale plots.}  
\label{fig:tvsk}
\end{figure}

In figure~\ref{fig:tvsk} (a) we  plotted the total evacuation time as a function of $k$ for the three updates. The evacuation time is monotonous and decreases with increasing $k$. In this congested two-dimensional situation the hybrid shuffle case is  closer to the random shuffle one, as expected. 
 
Now we apply an approximation, which is similar to what we did for $k=\infty$: all the sites $(x,y)$ with $ |x|+|y|=2  $   are assumed to be occupied at any integer time. For the calculation of hopping probabilities, we also use an approximation that  we take into account only the two choices, \textit{i.e.}   to stay at the same site and to move the neighbouring site  closer to the exit.  Then the probability of  particles' hops 
  $ (0,1) \to ( 0,0 ) $,  $ (-1,1) \to ( 0,1 ) $,  $ (0,2) \to ( 0,1 ) $ and   $ (1,1) \to (0,1 ) $ 
    (denoted by $p_0, p_1, p_2$ and $ p_3$,  respectively)   are given as  
\begin{equation}\label{eq:p_0123}
     p_0 = \frac{1}{1+e^{-k}}, \ 
     p_1 = p_3 =  \frac{e^{-k}}{e^{-k}+e^{-\sqrt 2 k}}, \ 
     p_2 =  \frac{e^{-k}}{e^{-k}+e^{- 2 k}}  .\ 
\end{equation}
Now we extend the master equation (\ref{eq:meqrsu}) to general $k$.  Since $(0,0)$ and $ (0,1 ) $ can be simultaneously empty,  we add $P^0_{\, 0} (t)$ to the probability vector: 
  \begin{equation}
\vec P(t  + 1 )  = M \vec P(t) , \ 
\vec P(t)  = 
\left( \begin{array}{l} P^0_{\, 0} (t)\\ P^0_{\, 1}(t) \\ P^1_{\, 0}(t)\\ P^1_{\, 1} (t)\end{array} \right)  .
\end{equation}
For the random-shuffle case, the elements of the transition probability matrix are given as 
\begin{eqnarray}
M = 
\left(\begin{array}{cccc}
   s_3 & s_3 & 0 & 0 \\
   0 & 0 & \frac{ 3+s_1+s_2+3 s_3 }{12} p_0 &
    \frac{  12+3 s_1+2 s_2+3 s_3 }{60} p_0 \\
   1-s_3 & 1-s_3 & 1-p_0 &  1 - \frac 1 2 p_0  \\
   0 & 0 &  \frac{  9 - s_1 - s_2 - 3 s_3 }{12} p_0  &  \frac{  18 - 3 s_1 - 2 s_2 - 3 s_3 }{60} p_0 
\end{array}\right) ,  
\end{eqnarray}
where 
$    s_1 = q_1 +  q_2  + q_3, \   
    s_2 = q_1 q_2 + q_1 q_3 + q_2 q_3, \   
    s_3 = q_1 q_2 q_3 , \ q_i = 1 -p_i  $.
Solving the balance equation $   \vec P = M \vec P $, we finally find  the outflow as 
\begin{eqnarray}
\label{eq:JR(k)}
  J_{\rm R}(k) =  P^0_{\, 1} +  P^1_{\, 1}   
 = \textstyle 
  \frac{ 120 p_0  (1-s_3 )     + p_0^2 (1-s_3 )  ( 9 + s_1 - s_2 - 9 s_3     ) }{
   120  ( 1- s_3 )  +  2 p_0   (42+3 s_1+2 s_2-24 s_3+2 s_1 s_3+3 s_2 s_3+12 s_3^2 ) 
    + p_0^2  ( 9+ s_1 - s_2 - 9 s_3 ) 
  }.  
\end{eqnarray}
The total evacuation time is then estimated by  $ T_{\rm R}(k) = N/ J_{\rm R}(k)  $.

For the frozen and hybrid cases, the correlation among phases prevents us from obtaining $M$ explicitly.  In figure~\ref{fig:tvsk}, however, we  \textit{luckily} observe that   the  plots $ T_{\rm H}(k) / T_{\rm H}(\infty) -1 $      and the   plots $ T_{\rm R}(k) / T_{\rm R}(\infty) -1 $    almost overlap. This means that the total evacuation time for the hybrid shuffle update  with finite $k$ is estimated as  $ T_{\rm H}(k) = T_{\rm H}(\infty)  J_{\rm R}(\infty)  / J_{\rm R}(k) $  by using (\ref{eq:JR(k)}). In the frozen case, when a particle of a platoon does not hop, another one may squeeze in and modify the structure of the platoons, hence a stronger effect of finite $k$ values.

  Finally we discuss on the crossover points  between the  high- and low-density regimes.  For finite $k$,  the inflow of the low-density regime  at time  $t$,  $J^\mathrm{in}(t)  $,  is not given by~(\ref{eq:Jin}). We rather determine it from single-particle simulations, using the approximation  
\begin{equation}
   J^\mathrm{in}(t) \approx
    N \times   \mathbb P (T_i = t ) , 
\end{equation}
where  $   \mathbb P (T_i = t )$ is  the probability of the evacuation time $T_i$  in each simulation run of the single-particle problem.   Then the crossover points are estimated from 
  $  \displaystyle \max_t  J^\mathrm{in}(t)  = J^\mathrm{out}    $ 
  \textit{i.e.}  
\begin{equation}
   N_c  \approx    J^\mathrm{out}   \big/   \max_t     \mathbb P (T_i = t )  
\end{equation} 
  with the outflows $ J^\mathrm{out}$   measured in the three updates with finite $k$.  For $k=3 $, the crossover points are numerically estimated to be    $N_c \approx 32.1, 48.4 $  and $34.2$, which provide good estimations,    see figure~\ref{fig:totalevactime} (c).

\section{Pedestrian evacuation}
\label{section:pedevac}

\subsection{Collective effects}

Some of the most impressive collective effects exhibited by pedestrian crowds  are found in evacuations,  see reference~\cite{schadschneider2009} for review. Countless human disasters have happened during mass events, where the motion of the crowd lead to stampedes~\cite{helbing2005, helbing_j_a2007}.

Modeling is useful to predict the crowd behavior, but also as a tool to understand the link between individual behavior and collective effects. In particular, interactions between individuals play a key role in evacuation processes.

Several types of interactions take place during evacuation processes. At \textit{very high} density,   the flow  at  a bottleneck can be completely stopped due to the formation of arches, in which the pedestrians are compacted and cannot move because of physical contact with their neighbours.

At low densities, pedestrians interact without physical contact through so-called social interactions, in order to avoid collisions while minimizing the inconvenience of deviating from their optimal  trajectory. A full characterization of these interactions is difficult, as they are non local, non isotropic, they include some anticipation, and they are non-additive when more than two pedestrians are involved.

Another way to characterize the interactions is to observe their consequences on the collective behavior and to infer from  their   most relevant characteristics of interactions.

\subsubsection{Obstacle in front of the exit}

One interesting feature is the role of an obstacle suitably located near the exit of a room. Some experiments have been performed, which compare the outflows from the room, with and without obstacle. When pedestrians are in a panic-like condition, \textit{i.e.} with strong physical contacts, the presence of a well located obstacle can surprisingly increase the outflow by about 30\%~\cite{helbing2005}. But for obvious security reasons, this effect has not been systematically confirmed\footnote{
A similar effect was observed for ants~\cite{shiwakoti2011} and for granular matter~\cite{zuriguel2011}, but the systems are too different to extrapolate to pedestrians.}.
 
In more ``normal'' conditions, the difference between outflows with and without obstacle are far less pronounced~\cite{helbing2005,yanagisawa2009,nishinari2010} (the outflow increase is estimated around 4\% in~\cite{yanagisawa2009}, 6\% in~\cite{nishinari2010}) and it should still be confirmed whether this weak increase is also observed in real life conditions. However, even if the outflows are more or less equal,  this is still surprising, as we could expect that an obstacle would \textit{decrease} the outflow (and indeed this is the case for ill-located or ill-sized obstacles~\cite{nishinari2010}).

\subsubsection{Faster-is-slower effect}

Another effect that was conjectured from numerical observations is the so-called faster-is-slower effect, namely the fact that a stronger will of pedestrians to go out may lead to longer evacuation times. An experiment of aircraft evacuation showed a trend to have larger evacuation times in competitive trials compared to non-competitive ones, but the result was strongly geometry dependent and opposite effects could also be observed~\cite{muir_b_m1996}. More recently, it was shown~\cite{garcimartin_2014,pastor_2015} that indeed a faster-is-slower effect could be obtained with humans, leading to an increase of evacuation times of typically $15$ to $20$\% under competitive egress, and related to physical contacts and the formation of clogs. A faster-is-slower effect can also be observed in some other systems that exhibit clogging, namely granular matter or sheep~\cite{pastor_2015}. By contrast, as ants do not clog but rather keep their density always more or less constant even under stress~\cite{parisi_2015}, no significant faster-is-slower effect was observed for ants~\cite{boari_j_p2013}, except in the trivial limit where the stress source has an impact on the physiology of the ants, and thus on their stepping behavior~\cite{parisi_2015}.   In this paper, we rather focus on \textit{normal} evacuation conditions of pedestrians, \textit{i.e.} with limited physical contacts, for which no faster-is-slower effect was ever evidenced. It is an open question to characterize the interactions between pedestrians in these more ordinary situations.

\subsubsection{1D and 2D outflows}

An experimental observation that may seem less spectacular but has strong implications for modeling, is that, if one compares the outflow of a single line and of two-dimensional flows, one finds that 2D outflows are similar or even sometimes smaller than 1D outflows\footnote{This remark applies for disordered random 2D flows. If pedestrians are pre-ordered into several incoming lines, higher flows can be achieved~\cite{yanagisawa2009,nishinari2010}, but this situation is quite special and we shall not consider it here. Still, it is an open question to know how to account for this phenomenon in simulations. Due to the discrete lattice, straight motions in directions not parallel to the lattice have to be decomposed into hops of different orientation. Besides, the exclusion rule is too crude to account for the subtle zipper effects that are favored with pre-ordered lines at the bottleneck~\cite{yanagisawa2009}.}. Indeed, in~\cite{yanagisawa2009}, two experiments are performed and  the reported decrease of the 2D outflow compared to the 1D outflow is respectively 4\% and 14\%.  

The fact that 2D outflows are not larger than 1D outflows is quite surprising.   In fact, in two-dimensional flows, more pedestrians arrive at the exit. This overfeeding effect should lead to higher flows\footnote{
Indeed, as we said in the previous footnote, this increase   is what happens with pre-ordered lines~\cite{yanagisawa2009,nishinari2010}.}. Actually this is what we observe when we compare the 1D and 2D outflows obtained in sections~\ref{subsection:updates} and~\ref{section:highd}~: Both for the frozen and random shuffle updates, the 2D outflow is larger than the one-dimensional one. This makes it more clear now why we have introduced the hybrid shuffle update defined in section~\ref{subsection:hybrid}. One sees from figure~\ref{fig:currentt}~(c) that for this update, the 2D and 1D outflows (both measured for the deterministic case $k=\infty$ to compare equivalent models) are similar.

The fact that 2D outflows are not larger than 1D outflows indicates that there are some interactions between pedestrians beyond the hard-core ones,  that compensate the overfeeding. As these interactions decrease the flow, they are often assimilated with friction. One must be careful that this notion of friction can refer to different types of interactions (with physical contacts or without in `normal' flows), with different consequences on the collective behavior as we explained above. The choice of the friction implemented in the simulations implicitly assumes whether physical contacts are considered or not and it is not clear yet whether a single model can account for all these different regimes.

\subsection{Updates}

\subsubsection{Parallel update}

Most cellular automata models for pedestrians use parallel update. This update has the particularity that conflicts occur: when two pedestrians have the same target site, the exclusion rule requires to select which one will actually hop. The need to include a conflict resolution procedure is in general considered as a drawback in simulations. However, in the case of pedestrian flows, it was suggested that these conflicts could have a physical relevance~\cite{kirchner2003a, kirchner_n_s2003,schadschneider_s2009}. Indeed, they occur preferentially in these converging areas where friction should occur. An easy way to tune the friction level is to  
  allow  with a certain probability none of the pedestrians involved to move. 
This clearly reduces the overall flow when the number
of conflicts increases.
 
\subsubsection{Shuffle updates and stepping}

One interesting feature of the frozen shuffle update is the possibility to interpret the phases ${\tau_i}$ as the phases in the stepping cycle of pedestrians. This allows in particular, for 1D deterministic flows, to map the cellular automaton dynamics in the free flow phase onto a continuous space and time dynamics~\cite{appert-rolland_c_h2011b}, reproducing the regular motion of free pedestrians. This also gives a physical interpretation to the priority given to one pedestrian when two pedestrians meet. The time step of the update, which plays the role of a reaction time, is then directly connected to the stepping period --  actually   it is known that pedestrians cannot turn/change velocity at any points of the stepping cycle but only at specific moments~\cite{olivier_c2007}. In the frame of this interpretation of the phases, it becomes natural to modify the phases in the jammed phase, as done by hybrid shuffle updates. Indeed there is no reason why a pedestrian should keep its stepping phase while being blocked by its predecessor.

It should be noted that cellular automata are rather mesoscopic models that do not aim to reproduce the precise microscopic dynamics but simply to give the correct average flows and densities at specific key locations. Still, it would be interesting to see if features coming from the stepping behavior of pedestrians could be retained and give interesting properties to the models, closer to what happens at the scale of individuals.

The hybrid shuffle update presented in this paper illustrates that indeed some interactions leading to flow reduction in the congested phase can be implemented through a partial renewal of the phases. The choice we have made is not unique (see~\cite{arita_c_a2014} for another one). One difficulty is to trigger friction preferentially in convergent flows -- and not in all high density situations\footnote{
For the parallel update, conflict occurrence automatically coincides with convergence of the flow.}. Still, we have not explored all the possibilities opened by hybrid updates. While in this paper the phase dynamics consists simply in a partial redrawing of the phases from a uniform distribution, one could also consider some evolution rules for the phases where the new phase value would depend for example on its previous value, on the phases of its neighbours, or possibly on the phase of the last visitor of the target cell.

\subsection{Fundamental diagrams}

 Flow-density    relations have been measured in several experiments.  However a consensus is difficult to reach, first, because the fundamental diagram depends on the geometry of the system, and second, because the precise method used to measure velocity and density may have a strong influence
 on the results~\cite{schadschneider_s2011,tordeux2015}. However, now that many data have been collected, one can nevertheless make a few observations.
 
A first set of experiments deals with one-dimensional flows, where pedestrians walk in a line without passing each other~\cite{seyfried2005,chattaraj_s_c2009a,jelic2012a}. If one would crudely approximate the fundamental diagram in this one-dimensional case by a triangular shape, the density for which the maximal flow  should be clearly smaller than half the maximal density~\cite{zhang2014}. This is not the case for the one-dimensional TASEP, for which the density of maximum flow is $\rho=1/2$ for random shuffle and parallel updates, and $\rho=2/3$ for frozen shuffle update.
 
It is known that one way to shift the maximum of the phase diagram towards smaller densities is to allow particles to jump more than one cell at each time step. Indeed this was shown  for the Nagel-Schreckenberg model for car traffic~\cite{nagel_s1992,chowdhury_s_s2000}, but also for the floor field model with parallel update~\cite{kirchner2004,schadschneider_s2009}.

In two-dimensional flows,  the variety of possible geometries (corridors, bottlenecks, intersections, etc) makes it more difficult to conclude. In some measurements~\cite{helbing_j_a2007,chraibi_s_s2010}, the density of maximal flow was found to be shifter towards higher densities than in 1D, while in some others not. Besides, pedestrians can have different walking directions, a feature which of course impacts the fundamental diagram. 

In the simulations, the level of discretization~\cite{kirchner2004}, or the details of the hopping rules~\cite{hao2010}, can also modify the shape of the fundamental diagram. Thus it is not possible to validate a model only on the basis of its update scheme, though the update scheme has an influence on the properties of the model. The whole set of characteristics of the model should be taken into account.

\subsection{Queue shapes}

\begin{figure}\begin{center}
    \includegraphics[width=0.6\textwidth]{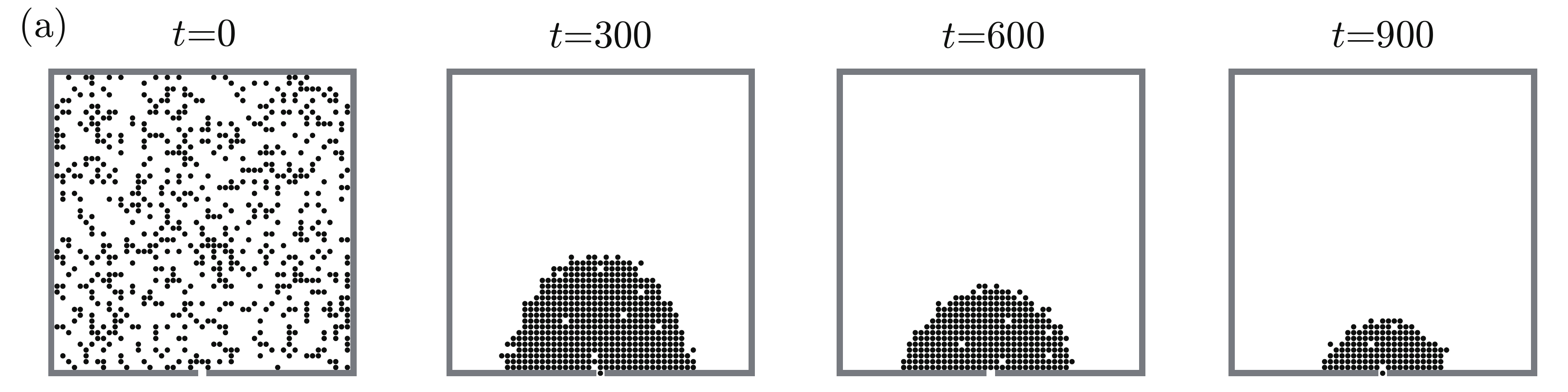} \\
     \includegraphics[width=0.6\textwidth]{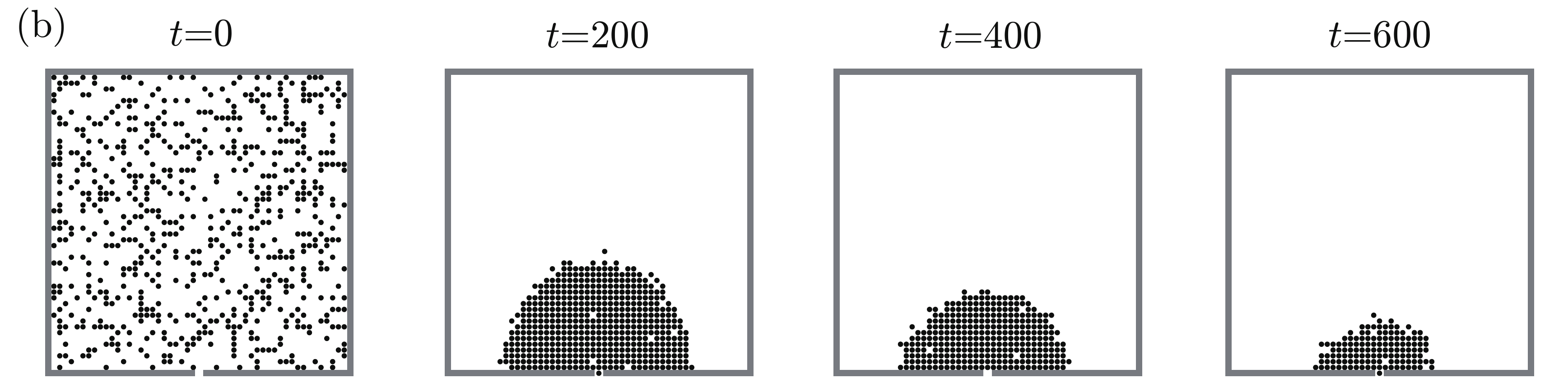} \\
     \includegraphics[width=0.6\textwidth]{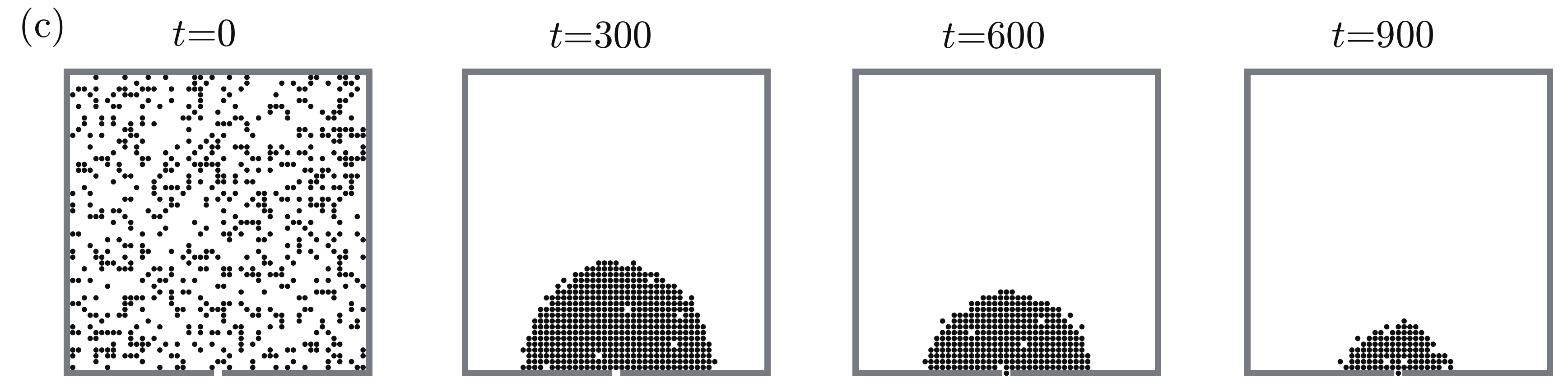} 
     \end{center}
\caption{Typical queue shapes for random shuffle (a), frozen shuffle (b) and hybrid shuffle (c) updates with parameters $L=51^2$, $N=(L^2-1)/4$ and $k = 5$. 
\label{fig:queueshape}
}
\end{figure}

We end this discussion by a few observations on the queue shapes. Though having a realistic queue shape in front of the bottleneck is not necessarily required as long as outflows are correct, it would be more satisfactory to approach  the shape of real queues. In figure~\ref{fig:queueshape}, we show the positions of particles at different moments of time,  as obtained with the model of this paper for $k=5$. The queues look like half-circles for the three updates considered. This is what is usually observed in simulations. In experiments, shapes like droplets have rather been observed (see figure 4 in \cite{daamen_h2003b}). This difference of queue shape between simulations and experiments was already mentioned in~\cite{schadschneider2009}. We have shown in reference~\,\cite{arita_c_a2014} a simulation snapshot exhibiting a queue with droplet shape,   as a result of some modifications of the rules, especially adding diagonal hopping. Indeed, it was shown in~\cite{tanimoto_h_t2011} how the queue shape can vary strongly (including very unrealistic shapes) depending on the allowed hops and their associated rates. Another way to decrease the density on each side of the exit is by penalizing sharp turns~\cite{yanagisawa2009}.

\section{Conclusion} \label{section:conclusion}

\begin{table}[t]
\begin{center}
\begin{tabular}{c|cccc}
\label{table:currents}
   & Random & Frozen & Hybrid\\
\hline 
 $J^\mathrm{1D}$    &    $1/2$& $2/3$ & $2/3$\\
 $J^\mathrm{2D}$   &   $  43/71  $ & $1$ & $ 0.64$\\
\end{tabular}
\end{center}
\caption{Values of the maximal currents for the random shuffle, frozen shuffle and hybrid shuffle updates in one and two dimensions (and for $k=\infty$). All the values of $J^\mathrm{1D}$  are exact in the limit of an infinite system.  The value of $J^\mathrm{2D}$ for frozen shuffle update   is also expected to be exact in the limit of an infinite system.    For random shuffle update, $ J^{\rm 2D} = \frac{43}{71} $ was led by an approximation,
 and the 2D value for hybrid shuffle update was measured numerically.
 \label{table}}\end{table}

In this article, we have investigated the properties of shuffle updates in two-dimensions through the study of a simple evacuation problem. A particular attention has been given to evacuation times and outflows, which we have studied both numerically and analytically.
 
In this study, we have considered a floor field model with only a static field. In general floor field models, a \textit{dynamic} floor field is also considered, which allows to transform some long range interactions between pedestrians into purely local ones~\cite{burstedde2001b}. Indeed our aim was not to design a full pedestrian model but to explore how interactions between pedestrians could be modified through the update scheme itself. More generally, it is an important issue to understand how to model properly inter-pedestrian interactions, as they determine in particular the outflow of evacuation problems.

We have used a unified picture to define the various shuffle updates, by associating a phase $\tau_i$ to each particle. Then updates differ only by the dynamics of these phases. For the random shuffle update,  all the phases  are redrawn in every time step.  For the frozen one,  phases are initially $(t=0)$ given to particles, and we keep them till the end of evacuation.  For the hybrid one, we partially redraw some of the  phases. The last  one interpolates between the frozen at low density and the random in congested configurations. Using the hybrid shuffle update the phases $\{\tau_i \}$ keep their interpretation as trackers of the walking cycle, and the dynamics of the phases in the high density phase simply corresponds to a loss of memory of the stepping cycle phase when pedestrians are forced to slow down or stop.

It is clear that many updating procedures that interpolate between frozen shuffle and random shuffle exist, and that the choice we have presented in this article is only a particular instance of them, that we have chosen to illustrate how model properties can be modified by including some phase dynamics. 
 We found that this test case  has similar 1D and 2D outflows. Table \ref{table} shows the flows for the shuffle updates, summarizing subsections~\ref{subsection:updates},  \ref{subsection:hybrid} and  \ref{subsection:maxcurrent}.

Other features could be included, in particular via the phase dynamics. For example, in reference~\cite{kretz_g_s2006} measurements of the time gap between two consecutive pedestrians are presented. In our model, phases allow to have an interpretation of the dynamics in continuous time, and hence it is possible to define time gaps. Attempts could be made to monitor the distribution of this observable, \textit{e.g.} by drawing the phase $\tau_i$ of a blocked particle in correlation with the phase of the blocking particle.

The idea of deriving an updating order from the stepping dynamics was even pushed further in the so-called  optimal steps model~\cite{seitz_k2012,seitz_k2014,dietrich2014}, for which not only the phase but also the duration of the stepping cycle and the length of the steps vary from one individual to another one. Though in the case of a constant step length the  optimal steps model  can also be applied to cellular automata models, it is more generally defined in a continuous space. It would be interesting to study the  influence of this update on outflows too.   

Improvement of pedestrian models would require a better understanding of flow-density relations. Beyond the fact that a better consensus should be reached on which fundamental diagram the models should verify, one must be aware that fundamental diagrams are supposed to be obtained in stationary regimes. In practice, in experiments, one realizes short-term and localized averages. In the extreme case where instantaneous individuals measurements are realized, one observes quite different flow-velocity relations. Indeed, the fluctuations around the mean behavior can be quite strong~\cite{jelic2012a}. It is an open question to determine whether real pedestrian flows -- including the important case of evacuations -- can be fully described by implementing stationary laws, or whether they could be dominated by transients, which can give quite different flow values, not necessarily well reproduced by models calibrated in the stationary regime. Another issue to improve cellular automata models would be to study more systematically the isotropy of their properties~\cite{schultz_f2010}.

\section*{Acknowledgements}

We thank Ludger Santen for useful discussions and Daniel R. Parisi for private communication.
We are grateful to an anonymous referee for critical reading of the manuscript and many valuable comments.


\vspace{5mm}

\begin{thebibliography}{10}


\bibitem{schadschneider2002a}
A.~Schadschneider, Cellular automaton approach to pedestrian dynamics - theory,
  in: M.~Schreckenberg, S.~Sharma (Eds.), Pedestrian and Evacuation Dynamics,
  Springer, 2002, pp. 75--85.

\bibitem{helbing2001b}
D.~Helbing, Traffic and related self-driven many-particle systems, Reviews of
  Modern Physics 73 (2001) 1067--1141.

\bibitem{chowdhury_s_s2000}
D.~Chowdhury, L.~Santen, A.~Schadschneider, Statistical physics of vehicular
  traffic and some related systems, Physics Reports 329 (2000) 199--329.

\bibitem{chou_m_z2011}
T.~Chou, K.~Mallick, R.~K.~P. Zia, Non-equilibrium statistical mechanics: from
  a paradigmatic model to biological transport, Reports on progress in physics
  74 (2011) 116601.

\bibitem{appert-rolland_e_s2015}
C.~Appert-Rolland, M.~Ebbinghaus, and L.~Santen,
Intracellular transport driven by cytoskeletal motors: General mechanisms and defects,
Physics Reports 593  (2015) 1--59. 


\bibitem{nagel_s1992}
K.~Nagel, M.~Schreckenberg, A cellular automaton model for freeway traffic, J.
  Phys. I 2 (1992) 2221--2229.

\bibitem{knospe2002b}
W.~Knospe, L.~Santen, A.~Schadschneider, M.~Schreckenberg, Single-vehicle data
  of highway traffic: microscopic description of traffic phases, Phys. Rev. E
  65 (2002) 056133.

\bibitem{hoogendoorn_d2003b}
S.~P. Hoogendoorn, W.~Daamen, Self-organization in pedestrian flow, in:
  S.~Hoogendoorn, S.~Luding, P.~Bovy, M.~Schreckenberg, D.~Wolf (Eds.), Traffic
  and Granular Flow '03, Springer-Verlag Berlin Heidelberg, 2005, pp. 373--382.

\bibitem{daamen_h2003a}
W.~Daamen, S.~Hoogendoorn, Qualitative results from pedestrian laboratory
  experiments, in: E.~Galea (Ed.), Pedestrian and Evacuation Dynamics, 2003,
  pp. 121--132.

\bibitem{moussaid2012}
M.~Moussa{\"i}d, E.~Guillot, M.~Moreau, J.~Fehrenbach, O.~Chabiron,
  S.~Lemercier, J.~Pettr\'e, C.~Appert-Rolland, P.~Degond, G.~Theraulaz,
  Traffic instabilities in self-organized pedestrian crowds, PLoS Computational
  Biology 8 (2012) 1002442.

\bibitem{plaue2011}
M.~Plaue, M.~Chen, G.~B{\"a}rwolff, H.~Schwandt, Trajectory extraction and
  density analysis of intersecting pedestrian flows from video recording, in:
  U.~Stilla, F.~Rottensteiner, H.~Mayer, B.~Jutzi, M.~Butenuth (Eds.),
  Photogrammetric Image Analysis, Vol. 6952, Springer Berlin Heidelberg, 2011,
  pp. 285--296.

\bibitem{zhang_s2014}
J.~Zhang, A.~Seyfried, Comparison of intersecting pedestrian flows based on
  experiments, Physica A-Stat. Mech. and its Applic. 405 (2014) 316--325.

\bibitem{barlovic1998}
R.~Barlovi\'c, L.~Santen, A.~Schadschneider, M.~Schreckenberg, Metastable
  states in cellular automata for traffic flow, Eur. Phys. J. B5 (1998)
  793--800.

\bibitem{appert_s2001}
C.~Appert, L.~Santen, Boundary induced phase transitions in driven lattice
  gases with meta-stable states, Phys. Rev. Lett. 86 (2001) 2498--2501.

\bibitem{burstedde2001a}
C.~Burstedde, A.~Kirchner, K.~Klauck, A.~Schadschneider, J.~Zittartz, Cellular
  automaton approach to pedestrian dynamics - applications, in:
  M.~Schreckenberg, S.~Sharma (Eds.), Pedestrian and Evacuation Dynamics,
  Springer, 2001, p.~87.

\bibitem{burstedde2001b}
C.~Burstedde, K.~Klauck, A.~Schadschneider, J.~Zittartz, Simulation of
  pedestrian dynamics using a 2-dimensional cellular automaton, Physica A 295
  (2001) 507--525.

\bibitem{kirchner2003a}
A.~Kirchner, H.~Kl{\"u}pfel, K.~Nishinari, A.~Schadschneider, M.~Schreckenberg,
  Simulation of competitive egress behaviour: comparison with aircraft
  evacuation data, Physica A 324 (2003) 689--697.


\bibitem{schadschneider_s2011}
A.~Schadschneider, A.~Seyfried, Empirical results for pedestrian dynamics and
  their implications for modeling, Networks and Heterogeneous Media 6 (2011)
  545--560.

\bibitem{cividini_a_h2013}
J.~Cividini, C.~Appert-Rolland, H.~Hilhorst, Diagonal patterns and chevron
  effect in intersecting traffic flows, Europhys. Lett. 102 (2013) 20002.

\bibitem{masuda_n_s2014}
T.~Masuda, K.~Nishinari, A.~Schadschneider, Critical bottleneck size for
  jamless particle flows in two dimensions, Phys. Rev. Lett. 112 (2014) 138701.

\bibitem{rajewsky1998}
N.~Rajewsky, L.~Santen, A.~Schadschneider, M.~Schreckenberg, The asymmetric
  exclusion process: Comparison of update procedures, J. Stat. Phys. 92 (1998)
  151--194.

\bibitem{kessel2002}
A.~Ke\ss el, H.~Kl{\"u}pfel, J.~Wahle, M.~Schreckenberg, Microscopic simulation
  of pedestrian crowd motion, in: ed.: M.~Schreckenberg, S.~D. Sharma (Eds.),
  Pedestrian and Evacuation Dynamics 2001, Berlin, Springer, 2002, pp.
  193--202.


\bibitem{appert-rolland_c_h2011a}
C.~Appert-Rolland, J.~Cividini, H.~Hilhorst, Frozen shuffle update for an
  asymmetric exclusion process on a ring, J. Stat. Mech. (2011) P07009.

\bibitem{wolki_s_s2006}
M.~W{\"o}lki, A.~Schadschneider, M.~Schreckenberg, Asymmetric exclusion
  processes with shuffled dynamics, J. Phys. A: Math. Gen. 39 (2006) 33--44.

\bibitem{wolki_s_s2007}
M.~W{\"o}lki, A.~Schadschneider, M.~Schreckenberg, Asymmetric exclusion
  processes with non-factorizing steady states, in: A.~Schadschneider,
  T.~Poschel, R.~Kuhne, M.~Schreckenberg, D.~Wolf (Eds.), Traffic and Granular
  Flow ' 05, 2007, pp. 473--479.

\bibitem{smith_w2007a}
D.~A. Smith, R.~E. Wilson, Dynamical pair approximation for cellular automata
  with shuffle update, J. Phys. A: Math. Theor. 40    
  (2007) 2651--2664.

\bibitem{smith_w2007b}
D.~A. Smith, R.~E. Wilson, Extension of cluster dynamics to cellular automata
  with shuffle update, in: A.~Schadschneider, T.~Poschel, R.~Kuhne,
  M.~Schreckenberg, D.~Wolf (Eds.), Traffic and Granular Flow ' 05, 2007, pp.
  467--472.

\bibitem{appert-rolland_c_h2011b}
C.~Appert-Rolland, J.~Cividini, H.~Hilhorst, Frozen shuffle update for a
  deterministic totally asymmetric simple exclusion process with open
  boundaries, J. Stat. Mech. (2011) P10013.

\bibitem{cividini_a_h2014b}
J.~Cividini, C.~Appert-Rolland, H.~Hilhorst, Frozen shuffle update in simple
  geometries : a first step to simulate pedestrians, in: U.~Weidmann,
  U.~Kirsch, M.~Schreckenberg (Eds.), Pedestrian and Evacuation Dynamics 2012,
  Springer, Heidelberg, 2014, pp. 683--689.

\bibitem{kirchner_n_s2003}
A.~Kirchner, K.~Nishinari, A.~Schadschneider, Friction effects and clogging in
  a cellular automaton model for pedestrian dynamics, Phys. Rev. E 67 (2003)
  056122.

\bibitem{yanagisawa2009}
D.~Yanagisawa, A.~Kimura, A.~Tomoeda, R.~Nishi, Y.~Suma, K.~Ohtsuka,
  K.~Nishinari, Introduction of frictional and turning function for pedestrian
  outflow with an obstacle, Phys. Rev. E 80 (2009) 036110.

\bibitem{nishinari2010}
K.~Nishinari, Y.~Suma, D.~Yanagisawa, A.~Tomoeda, A.~Kimura, R.~Nishi, Toward
  smooth movement of crowds, in: ed.: W.W.F.~Klingsch, A.~Schadschneider,
  M.~Schreckenberg (Eds.), Pedestrian and Evacuation Dynamics 2008, Berlin,
  Springer, 2010, pp. 293--308.

\bibitem{arita_c_a2014}
C.~Arita, J.~Cividini, C.~Appert-Rolland, Shuffle updates in an evacuation
  problem, in: W.~Daamen, D.~Duives, S.~Hoogendoorn (Eds.), Pedestrian and
  Evacuation Dynamics 2014, Vol.~2 of Transportation Research Procedia,
  Elsevier, 2014, pp. 309--317.

\bibitem{schadschneider2009}
A.~Schadschneider, W.~Klingsch, H.~Kl{\"u}pfel, T.~Kretz, C.~Rogsch,
  A.~Seyfried, Evacuation dynamics: Empirical results, modeling and
  applications, in: B.~Meyers (Ed.), Encyclopedia of Complexity and System
  Science, Springer, Berlin, 2009, pp. 3142--3176.

\bibitem{helbing2005}
D.~Helbing, L.~Buzna, A.~Johansson, T.~Werner, Self-organized pedestrian crowd
  dynamics: Experiments, simulations, and design solutions, Transportation
  Science 39 (2005) 1--24.

\bibitem{helbing_j_a2007}
D.~Helbing, A.~Johansson, H.~Z. Al-Abideen, The dynamics of crowd disasters: An
  empirical study, Phys. Rev. E 75 (2007) 046109.

\bibitem{shiwakoti2011}
N.~Shiwakoti, M.~Sarvi, G.~Rose, M.~Burd, Animal dynamics based approach for
  modeling pedestrian crowd egress under panic conditions, Transportation
  Research B 45 (2011) 1433--1449.

\bibitem{zuriguel2011}
I.~Zuriguel, A.~Janda, A.~Garcimart\'in, C.~Lozano, R.~Ar\'evalo, D.~Maza,
 Silo  clogging reduction by the presence of an obstacle, Phys. Rev. Lett. 107
  (2011) 278001.

\bibitem{muir_b_m1996}
H.~C. Muir, D.~Bottomley, C.~Marrison, Effects of motivation and cabin
  configuration on emergency aircraft evacuation behavior and rates of egress,
  The International Journal of Aviation Psychology 6 (1996) 57--77.

\bibitem{garcimartin_2014}
A.~Garcimart\'in, I.~Zuriguel, J.~M.~Pastor, C.~Mart\'in-G\'omez, D.~R.~Parisi,
  Experimental evidence of the ``faster is slower'' effect, Transp. Res.
  Procedia 2 (2014) 760--767.

\bibitem{pastor_2015}
  J.~M.~Pastor et al, 
  Experimental proof of faster-is-slower in multi-particle systems flowing through bottlenecks, 
  arXiv: 1507.05110 (2015)

\bibitem{parisi_2015}
  D.~R.~Parisi, S.~Soria, R.~Josens, 
  Faster-is-slower effect in escaping ants revisited: Ants do not behave like humans, 
 Safety Science 72 (2015) 274--282.

\bibitem{boari_j_p2013}
S.~Boari, R.~Josens, D.~Parisi, Efficient egress of escaping ants stressed with
  temperature, PLOS One 8 (2013) e81082.

\bibitem{schadschneider_s2009}
A.~Schadschneider, A.~Seyfried, Validation of CA models of pedestrian
  dynamics with fundamental diagrams, Cybernetics and systems 40 (2009)
  367--389.

\bibitem{olivier_c2007}
A.-H. Olivier, A.~Cr\'etual, Velocity/curvature relations along a single turn
  in human locomotion, Neuroscience Letters 412 (2007) 148--153.

\bibitem{tordeux2015}
A.~Tordeux, J.~Zhang, B.~Steffen, A.~Seyfried, Quantitative comparison of
  estimations for the density within pedestrian streams, J. Stat. Mech. (2015)
  P06030.

\bibitem{seyfried2005}
A.~Seyfried, B.~Steffen, W.~Klingsch, M.~Boltes, The fundamental diagram of
  pedestrian movement revisited, J. Stat. Mech. (2005) P10002.

\bibitem{chattaraj_s_c2009a}
U.~Chattaraj, A.~Seyfried, P.~Chakroborty, Comparison of pedestrian fundamental
  diagram across cultures, Advances in Complex Systems 12 (2009) 393--405.

\bibitem{jelic2012a}
A.~Jeli\'c, C.~Appert-Rolland, S.~Lemercier, J.~Pettr\'e, Properties of
  pedestrians walking in line -- fundamental diagrams, Phys. Rev. E 85 (2012)
  036111.

\bibitem{zhang2014}
J.~Zhang, W.~Mehner, S.~Holl, M.~Boltes, E.~Andresen, A.~Schadschneider,
  A.~Seyfried, Universal flow-density relation of single-file bicycle,
  pedestrian and car motion, Phys. Lett. A 378 (2014) 3274--3277.

\bibitem{kirchner2004}
A.~Kirchner, H.~Kl{\"u}pfel, K.~Nishinari, A.~Schadschneider, M.~Schreckenberg,
  Discretization effects and the influence of walking speed in cellular
  automata models for pedestrian dynamics, J. Stat. Mech. (2004) P10011.


\bibitem{chraibi_s_s2010}
M.~Chraibi, A.~Seyfried, A.~Schadschneider, Generalized centrifugal-force model
  for pedestrian dynamics, Phys. Rev. E 82 (2010) 046111.

\bibitem{hao2010}
Q.~Hao, M.~Hu, X.~Cheng, W.~Song, R.~Jiang, Q.~Wu, Pedestrian flow in a lattice
  gas model with parallel update, Phys. Rev. E 82 (2010) 026113.

\bibitem{daamen_h2003b}
W.~Daamen, S.~Hoogendoorn, Experimental research of pedestrian walking
  behavior, Transportation Research Record 1828 (2003) 20--30.

\bibitem{tanimoto_h_t2011}
J.~Tanimoto, A.~Hagishima, Y.~Tanaka, Study of bottleneck effect at an
  emergency evacuation exit using cellular automata model, mean field
  approximation analysis, and game theory, Physica A 389 (2011) 5611--5618.

\bibitem{kretz_g_s2006}
T.~Kretz, A.~Gr{\"u}nebohm, M.~Schreckenberg, Experimental study of pedestrian
  flow through a bottleneck, J. Stat. Mech. (2006) P10014.

\bibitem{seitz_k2012}
M.~Seitz, G.~K{\"o}ster, Natural discretization of pedestrian movement in
  continuous space, Phys. Rev. E (2012) 046108.

\bibitem{seitz_k2014}
M.~Seitz, G.~K{\"o}ster, How update schemes influence crowd simulations, J.
  Stat. Mech. (2014) P07002.

\bibitem{dietrich2014}
F.~Dietrich, G.~K{\"o}ster, M.~Seitz, I.~von Sivers, Bridging the gap: From
  cellular automata to differential equation models for pedestrian dynamics,
  Journal of Computational Science 5 (2014) 841--846.

\bibitem{schultz_f2010}
M.~Schultz, H.~Fricke, Stochastic transition model for discrete agent
  movements, Lecture Notes in Computer Science 6350 (2010) 506--512.

\end{thebibliography}
\end{document}